\documentclass[11pt]{article}

\usepackage{xcolor}
\usepackage{hyperref}
\hypersetup{
    colorlinks,
    linkcolor={red!50!black},
    citecolor={blue!50!black},
    urlcolor={blue!80!black}
}
\usepackage[authoryear,round]{natbib} 

\usepackage[utf8]{inputenc}
\usepackage[T1]{fontenc}
\usepackage{amsmath,amssymb}
\usepackage{graphicx}
\usepackage[margin=2.5cm]{geometry}
\usepackage{booktabs}
\usepackage{siunitx}
\usepackage{hyperref}
\usepackage{natbib}
\usepackage{caption}
\usepackage{subcaption}
\usepackage{xcolor}
\usepackage{float}
\usepackage[section]{placeins}

\newcommand{\fitwidth}[1]{\resizebox{\ifdim\width>\linewidth\linewidth\else\width\fi}{!}{#1}}

\newcommand{\dpxMin}{11.5}
\newcommand{\dpxMax}{24.3}
\newcommand{\dpxMedian}{11.7}
\newcommand{\nDpxFlows}{15}
\newcommand{\dpxLargeMin}{11.5}
\newcommand{\dpxLargeMax}{24.3}
\newcommand{\errStep}{6}
\newcommand{\errD}{0.81}
\newcommand{\errFrame}{35}
\newcommand{\errSrcFrame}{210}
\newcommand{\errBTHota}{0.963}
\newcommand{\errBTAssA}{0.928}
\newcommand{\errBTFone}{0.9998}
\newcommand{\errBTSwapCell}{284}
\newcommand{\errBTMed}{0.038}
\newcommand{\errBTMedLo}{0.033}
\newcommand{\errBTMedHi}{0.040}
\newcommand{\errBTSwaps}{11}
\newcommand{\errBTSwapWin}{212}
\newcommand{\errTaHota}{0.929}
\newcommand{\errTaAssA}{0.863}
\newcommand{\errTaFone}{1.0000}
\newcommand{\errTaSwapCell}{412}
\newcommand{\errTaMed}{0.020}
\newcommand{\errTaMedLo}{0.018}
\newcommand{\errTaMedHi}{0.024}
\newcommand{\errTaSwaps}{6}
\newcommand{\errTaSwapWin}{144}
\newcommand{\errClip}{0.098}
\newcommand{\nCells}{59}
\newcommand{\nFlows}{15}
\newcommand{\nBaselines}{12}
\newcommand{\dMin}{0.006}
\newcommand{\dMax}{9.0}
\newcommand{\nBTLeadSmooth}{6}
\newcommand{\dCollOne}{0.14}
\newcommand{\collBTHotaOne}{1.000}
\newcommand{\collBTPtrOne}{1.000}
\newcommand{\collGothHotaOne}{0.992}
\newcommand{\collGothPtrOne}{0.999}
\newcommand{\collDeepOCHotaOne}{0.994}
\newcommand{\collDeepOCPtrOne}{1.000}
\newcommand{\collTpNVPHotaOne}{0.999}
\newcommand{\collTpNVPPtrOne}{1.000}
\newcommand{\collByteTrackHotaOne}{0.999}
\newcommand{\collByteTrackPtrOne}{1.000}
\newcommand{\dCollFour}{0.58}
\newcommand{\collBTHotaFour}{0.995}
\newcommand{\collBTPtrFour}{1.000}
\newcommand{\collGothHotaFour}{0.877}
\newcommand{\collGothPtrFour}{0.987}
\newcommand{\collDeepOCHotaFour}{0.772}
\newcommand{\collDeepOCPtrFour}{0.906}
\newcommand{\collTpNVPHotaFour}{0.787}
\newcommand{\collTpNVPPtrFour}{0.755}
\newcommand{\collByteTrackHotaFour}{0.652}
\newcommand{\collByteTrackPtrFour}{0.655}
\newcommand{\dCollEight}{1.22}
\newcommand{\collBTHotaEight}{0.904}
\newcommand{\collBTPtrEight}{0.974}
\newcommand{\collVFourHotaEight}{0.715}
\newcommand{\collGothHotaEight}{0.610}
\newcommand{\collGothPtrEight}{0.895}
\newcommand{\collDeepOCHotaEight}{0.388}
\newcommand{\collDeepOCPtrEight}{0.507}
\newcommand{\collTpNVPHotaEight}{0.654}
\newcommand{\collTpNVPPtrEight}{0.641}
\newcommand{\collByteTrackHotaEight}{0.363}
\newcommand{\collByteTrackPtrEight}{0.424}
\newcommand{\collBTMarginFieldEight}{+0.137}
\newcommand{\collBTMarginGothEight}{+0.294}
\newcommand{\collBTPtrPctEight}{97.4}
\newcommand{\collBTMaxScanMargin}{+0.14}
\newcommand{\dChuteA}{0.62}
\newcommand{\regChuteATpNVPHota}{0.541}
\newcommand{\regChuteATpNVPPtr}{0.992}
\newcommand{\regChuteABTHota}{0.540}
\newcommand{\regChuteABTPtr}{0.989}
\newcommand{\dChuteB}{0.84}
\newcommand{\regChuteBTpNVPHota}{0.359}
\newcommand{\regChuteBTpNVPPtr}{0.720}
\newcommand{\regChuteBBTHota}{0.466}
\newcommand{\regChuteBBTPtr}{0.982}
\newcommand{\dChuteC}{1.13}
\newcommand{\regChuteCTpNVPHota}{0.295}
\newcommand{\regChuteCTpNVPPtr}{0.656}
\newcommand{\regChuteCBTHota}{0.405}
\newcommand{\regChuteCBTPtr}{0.980}
\newcommand{\dLargeA}{1.15}
\newcommand{\regLargeATpNVPHota}{0.757}
\newcommand{\regLargeATpNVPPtr}{0.820}
\newcommand{\regLargeABTHota}{0.741}
\newcommand{\regLargeABTPtr}{0.785}
\newcommand{\dLargeB}{1.16}
\newcommand{\regLargeBTpNVPHota}{0.750}
\newcommand{\regLargeBTpNVPPtr}{0.833}
\newcommand{\regLargeBBTHota}{0.778}
\newcommand{\regLargeBBTPtr}{0.894}
\newcommand{\agChuteBBTHota}{0.466}
\newcommand{\agChuteCBTHota}{0.405}
\newcommand{\agHopperBTHota}{0.440}
\newcommand{\agBTPtrMin}{0.974}
\newcommand{\agBTPtrMax}{0.982}
\newcommand{\agTpPtrMin}{0.656}
\newcommand{\agTpPtrMax}{0.777}
\newcommand{\dCrowd}{0.56}
\newcommand{\swapTpKThreeEight}{47}
\newcommand{\swapBTKThreeEight}{0}
\newcommand{\occlDeepOC}{0.952}
\newcommand{\occlBT}{0.941}
\newcommand{\extDeepOCMin}{0.28}
\newcommand{\extDeepOCMax}{0.43}
\newcommand{\extGeoMax}{0.14}
\newcommand{\extBTMaxThree}{0.28}
\newcommand{\dMTwoThree}{1.9}
\newcommand{\dCollFirstLead}{0.7}
\newcommand{\dMTwoFirstLead}{1.2}
\newcommand{\dMTwoLastLead}{1.9}
\newcommand{\dMSixFirstLead}{0.7}
\newcommand{\dResAFirstLead}{1.2}
\newcommand{\smBTLeadDMin}{0.62}
\newcommand{\smBTMaxLead}{+0.102}
\newcommand{\smBTMaxDeficit}{0.025}
\newcommand{\smBTBehindDMax}{1.15}
\newcommand{\resBFourKEightBT}{0.741}
\newcommand{\resBFourKEightVFour}{0.336}
\newcommand{\resBFourKEightBest}{0.757}
\newcommand{\nLargeCells}{16}
\newcommand{\ablShipMean}{0.6515}
\newcommand{\ablBMean}{0.6736}
\newcommand{\ablBTMean}{0.6967}
\newcommand{\ablNCells}{53}
\newcommand{\ablBTImproved}{29}
\newcommand{\ablBTWorse}{0}
\newcommand{\ablResBFourKEightShip}{0.320}
\newcommand{\ablResBFourKEightB}{0.439}
\newcommand{\ablResBFourKEightBT}{0.740}
\newcommand{\ablResBFourKEightShipSwap}{18{,}250}
\newcommand{\ablResBFourKEightBSwap}{16{,}014}
\newcommand{\ablResBFourKEightBTSwap}{5{,}868}
\newcommand{\ablSmEightBGainMin}{+0.02}
\newcommand{\ablSmEightBGainMax}{+0.12}
\newcommand{\ablSmEightTGainMin}{+0.07}
\newcommand{\ablSmEightTGainMax}{+0.30}
\newcommand{\tqfVFourColGateLow}{23.8}
\newcommand{\tqfVFourAssignedLow}{70}
\newcommand{\tqfVFourColGateMid}{47.5}
\newcommand{\tqfVFourAssignedMid}{27}
\newcommand{\tqfVFourAssignedHigh}{6}
\newcommand{\tqfVFourBeyondDistHigh}{87}
\newcommand{\tqfBTColGateLow}{0.3}
\newcommand{\tqfBTAssignedLow}{98}
\newcommand{\tqfBTColGateMid}{0.4}
\newcommand{\tqfBTAssignedMid}{96}
\newcommand{\tqfBTAssignedHigh}{59}
\newcommand{\tqfBTBeyondDistHigh}{86}
\newcommand{\ablMTwoFourShip}{0.246}
\newcommand{\ablMTwoFourB}{0.381}
\newcommand{\ablMTwoFourBT}{0.431}
\newcommand{\hoNCells}{6}
\newcommand{\hoNCtx}{8}
\newcommand{\hoNPass}{6}
\newcommand{\hoBTMean}{0.7467}
\newcommand{\hoShipMean}{0.6525}
\newcommand{\hoVerdict}{confirmed}
\newcommand{\hoHOneOneBTHota}{0.867}
\newcommand{\hoHOneOneBTPtr}{0.989}
\newcommand{\hoHOneOneVFourHota}{0.823}
\newcommand{\hoHOneOneVFourPtr}{0.985}
\newcommand{\hoHOneOneCtxBestHota}{0.827}
\newcommand{\hoHOneOneCtxBestArm}{btrack}
\newcommand{\hoHOneFourBTHota}{0.586}
\newcommand{\hoHOneFourBTPtr}{0.826}
\newcommand{\hoHOneFourVFourHota}{0.495}
\newcommand{\hoHOneFourVFourPtr}{0.769}
\newcommand{\hoHOneFourCtxBestHota}{0.516}
\newcommand{\hoHOneFourCtxBestArm}{LapTrack}
\newcommand{\hoHOneEightBTHota}{0.493}
\newcommand{\hoHOneEightBTPtr}{0.655}
\newcommand{\hoHOneEightVFourHota}{0.362}
\newcommand{\hoHOneEightVFourPtr}{0.555}
\newcommand{\hoHOneEightCtxBestHota}{0.389}
\newcommand{\hoHOneEightCtxBestArm}{LapTrack}
\newcommand{\hoHTwoOneBTHota}{0.880}
\newcommand{\hoHTwoOneBTPtr}{0.996}
\newcommand{\hoHTwoOneVFourHota}{0.880}
\newcommand{\hoHTwoOneVFourPtr}{0.996}
\newcommand{\hoHTwoOneCtxBestHota}{0.882}
\newcommand{\hoHTwoOneCtxBestArm}{trackpy}
\newcommand{\hoHTwoFourBTHota}{0.874}
\newcommand{\hoHTwoFourBTPtr}{0.985}
\newcommand{\hoHTwoFourVFourHota}{0.870}
\newcommand{\hoHTwoFourVFourPtr}{0.982}
\newcommand{\hoHTwoFourCtxBestHota}{0.862}
\newcommand{\hoHTwoFourCtxBestArm}{trackpy-NVP}
\newcommand{\hoHTwoEightBTHota}{0.780}
\newcommand{\hoHTwoEightBTPtr}{0.931}
\newcommand{\hoHTwoEightVFourHota}{0.484}
\newcommand{\hoHTwoEightVFourPtr}{0.726}
\newcommand{\hoHTwoEightCtxBestHota}{0.650}
\newcommand{\hoHTwoEightCtxBestArm}{trackpy-NVP}
\newcommand{\lrnTaN}{59}
\newcommand{\lrnTaMean}{0.524}
\newcommand{\lrnTaBTMean}{0.668}
\newcommand{\lrnTaDevN}{4}
\newcommand{\lrnTaDevCells}{RES\_A\_4K s1, RES\_A\_4K s4, RES\_A\_4K s8, K05 s1}
\newcommand{\lrnCoNoBirthN}{53}
\newcommand{\lrnCoN}{42}
\newcommand{\lrnCoFailN}{11}
\newcommand{\lrnCoMean}{0.226}
\newcommand{\lrnCoBTMean}{0.714}
\newcommand{\lrnCoTaMean}{0.519}
\newcommand{\collTaHotaOne}{0.999}
\newcommand{\collTaPtrOne}{1.000}
\newcommand{\collCoHotaOne}{0.457}
\newcommand{\collTaHotaFour}{0.992}
\newcommand{\collTaPtrFour}{0.999}
\newcommand{\collCoHotaFour}{0.490}
\newcommand{\collTaHotaEight}{0.767}
\newcommand{\collTaPtrEight}{0.773}
\newcommand{\nCollScanTie}{4}
\newcommand{\collCoMaxresHotaOne}{0.809}
\newcommand{\collCoMaxresHotaFour}{0.861}
\newcommand{\coRegCollN}{7}
\newcommand{\coRegCollFail}{0}
\newcommand{\coRegCollCo}{0.483}
\newcommand{\coRegCollTa}{0.949}
\newcommand{\coRegCollBT}{0.978}
\newcommand{\coRegAgN}{3}
\newcommand{\coRegAgFail}{2}
\newcommand{\coRegAgCo}{0.097}
\newcommand{\coRegAgTa}{0.396}
\newcommand{\coRegAgBT}{0.506}
\newcommand{\coRegSmoothN}{23}
\newcommand{\coRegSmoothFail}{0}
\newcommand{\coRegSmoothCo}{0.209}
\newcommand{\coRegSmoothTa}{0.522}
\newcommand{\coRegSmoothBT}{0.818}
\newcommand{\coRegBedsN}{0}
\newcommand{\coRegBedsFail}{1}
\newcommand{\coRegExtN}{6}
\newcommand{\coRegExtFail}{5}
\newcommand{\coRegExtCo}{0.098}
\newcommand{\coRegExtTa}{0.118}
\newcommand{\coRegExtBT}{0.262}
\newcommand{\coRegDrumN}{3}
\newcommand{\coRegDrumFail}{3}
\newcommand{\coRegDrumCo}{0.146}
\newcommand{\coRegDrumTa}{0.421}
\newcommand{\coRegDrumBT}{0.412}
\newcommand{\coPartMThreeOneHota}{0.162}
\newcommand{\coPartMThreeOneEnd}{395}
\newcommand{\coPartMThreeOneD}{0.84}
\newcommand{\coPartMThreeOneBT}{0.466}
\newcommand{\coPartMThreeFourHota}{0.087}
\newcommand{\coPartMThreeFourEnd}{98}
\newcommand{\coPartMThreeFourD}{3.38}
\newcommand{\coPartMThreeFourBT}{0.085}
\newcommand{\coPartMThreeEightHota}{0.118}
\newcommand{\coPartMThreeEightEnd}{45}
\newcommand{\coPartMThreeEightD}{6.76}
\newcommand{\coPartMThreeEightBT}{0.116}
\newcommand{\coPartMFourOneHota}{0.276}
\newcommand{\coPartMFourOneEnd}{264}
\newcommand{\coPartMFourOneD}{1.13}
\newcommand{\coPartMFourOneBT}{0.405}
\newcommand{\coPartMFourFourHota}{0.123}
\newcommand{\coPartMFourFourEnd}{73}
\newcommand{\coPartMFourFourD}{4.54}
\newcommand{\coPartMFourFourBT}{0.086}
\newcommand{\coPartMFourEightHota}{0.049}
\newcommand{\coPartMFourEightEnd}{24}
\newcommand{\coPartMFourEightD}{9.03}
\newcommand{\coPartMFourEightBT}{0.117}
\newcommand{\coPartMFiveEightHota}{0.076}
\newcommand{\coPartMFiveEightEnd}{48}
\newcommand{\coPartMFiveEightD}{8.24}
\newcommand{\coPartMFiveEightBT}{0.126}
\newcommand{\coPartKFiveOneHota}{0.663}
\newcommand{\coPartKFiveOneEnd}{304}
\newcommand{\coPartKFiveOneD}{0.60}
\newcommand{\coPartKFiveOneBT}{0.954}
\newcommand{\coPartMOneDrumOneHota}{0.073}
\newcommand{\coPartMOneDrumOneEnd}{393}
\newcommand{\coPartMOneDrumOneD}{0.53}
\newcommand{\coPartMOneDrumOneBT}{0.640}
\newcommand{\coPartDrumSlowFourHota}{0.143}
\newcommand{\coPartDrumSlowFourEnd}{73}
\newcommand{\coPartDrumSlowFourD}{0.88}
\newcommand{\coPartDrumSlowFourBT}{0.461}
\newcommand{\coPartDrumSlowEightHota}{0.181}
\newcommand{\coPartDrumSlowEightEnd}{19}
\newcommand{\coPartDrumSlowEightD}{1.44}
\newcommand{\coPartDrumSlowEightBT}{0.168}
\newcommand{\tieShipExact}{28}
\newcommand{\tieShipWithin}{44}
\newcommand{\tieShipBeyond}{9}
\newcommand{\tieShipMaxHota}{0.0192}
\newcommand{\tieShipMaxAny}{0.0255}
\newcommand{\tieShipBeyondCells}{M02 s3 ($+0.0152$/$+0.0227$), M02 s4 ($-0.0192$/$-0.0255$), M02 s5 ($+0.0026$/$+0.0068$), M06 s5 ($+0.0061$/$+0.0024$), M06 s6 ($-0.0100$/$-0.0130$), M06 s8 ($+0.0018$/$-0.0057$), RES\_A\_1080 s4 ($-0.0027$/$-0.0031$), RES\_A\_1080 s8 ($+0.0021$/$+0.0035$), RES\_B\_4K s8 ($-0.0023$/$-0.0036$)}
\newcommand{\tieBTExact}{35}
\newcommand{\tieBTWithin}{50}
\newcommand{\tieBTBeyond}{3}
\newcommand{\tieBTMaxHota}{0.0085}
\newcommand{\tieBTMaxAny}{0.0147}
\newcommand{\tieBTBeyondCells}{M02 s4 ($-0.0085$/$-0.0115$), M02 s5 ($-0.0044$/$-0.0147$), M06 s8 ($-0.0019$/$-0.0023$)}
\newcommand{\mTwoThreeBTFieldMargin}{+0.048}
\newcommand{\mTwoFourBTFieldMargin}{-0.011}
\newcommand{\eligNonFloor}{48}
\newcommand{\eligViolations}{0}
\newcommand{\eligVerdict}{eligible}
\newcommand{\hoTzNPass}{6}
\newcommand{\hoTzBTMean}{0.7477}
\newcommand{\hoTzShipMean}{0.6524}
\newcommand{\hoTzVerdict}{confirmed}
\newcommand{\fixShipExact}{50}
\newcommand{\fixShipWithin}{51}
\newcommand{\fixShipBeyondCells}{RES\_B\_4K s4 ($-0.0002$/$-0.0045$), RES\_B\_4K s8 ($+0.0177$/$+0.0046$)}
\newcommand{\fixBTExact}{41}
\newcommand{\fixBTWithin}{51}
\newcommand{\fixBTBeyondCells}{RES\_A\_1080 s6 ($+0.0005$/$+0.0021$), RES\_B\_4K s8 ($+0.0011$/$+0.0075$)}
\newcommand{\fixHeldMaxAny}{0.0000}
\newcommand{\senNCells}{30}
\newcommand{\senCoastHalfName}{coast EMA $\times 0.5$}
\newcommand{\senCoastHalfValue}{$1.0\rightarrow 0.5$}
\newcommand{\senCoastHalfMeanHota}{$-0.0002$}
\newcommand{\senCoastHalfMeanPtr}{$-0.0012$}
\newcommand{\senCoastHalfMeanAbsHota}{0.0019}
\newcommand{\senCoastHalfMaxHota}{0.0146}
\newcommand{\senCoastHalfMaxHotaCell}{RES\_B\_4K s8}
\newcommand{\senCoastHalfMaxPtr}{0.0137}
\newcommand{\senCoastHalfMaxPtrCell}{RES\_B\_1080 s8}
\newcommand{\senCoastHalfStandHota}{2}
\newcommand{\senCoastHalfStandPtr}{0}
\newcommand{\senExitHalfName}{exit margin $\times 0.5$}
\newcommand{\senExitHalfValue}{$1.0\rightarrow 0.5$}
\newcommand{\senExitHalfMeanHota}{$+0.0001$}
\newcommand{\senExitHalfMeanPtr}{$0.0000$}
\newcommand{\senExitHalfMeanAbsHota}{0.0004}
\newcommand{\senExitHalfMaxHota}{0.0041}
\newcommand{\senExitHalfMaxHotaCell}{RES\_A\_1080 s6}
\newcommand{\senExitHalfMaxPtr}{0.0079}
\newcommand{\senExitHalfMaxPtrCell}{RES\_A\_1080 s8}
\newcommand{\senExitHalfStandHota}{1}
\newcommand{\senExitHalfStandPtr}{1}
\newcommand{\senExitTwoName}{exit margin $\times 2$}
\newcommand{\senExitTwoValue}{$1.0\rightarrow 2.0$}
\newcommand{\senExitTwoMeanHota}{$+0.0001$}
\newcommand{\senExitTwoMeanPtr}{$0.0000$}
\newcommand{\senExitTwoMeanAbsHota}{0.0003}
\newcommand{\senExitTwoMaxHota}{0.0022}
\newcommand{\senExitTwoMaxHotaCell}{RES\_B\_1080 s4}
\newcommand{\senExitTwoMaxPtr}{0.0053}
\newcommand{\senExitTwoMaxPtrCell}{RES\_B\_4K s4}
\newcommand{\senExitTwoStandHota}{2}
\newcommand{\senExitTwoStandPtr}{1}
\newcommand{\senColHalfName}{colour EMA $\times 0.5$}
\newcommand{\senColHalfValue}{$0.3\rightarrow 0.15$}
\newcommand{\senColHalfMeanHota}{$-0.0023$}
\newcommand{\senColHalfMeanPtr}{$-0.0038$}
\newcommand{\senColHalfMeanAbsHota}{0.0029}
\newcommand{\senColHalfMaxHota}{0.0408}
\newcommand{\senColHalfMaxHotaCell}{RES\_B\_4K s8}
\newcommand{\senColHalfMaxPtr}{0.0713}
\newcommand{\senColHalfMaxPtrCell}{RES\_B\_4K s8}
\newcommand{\senColHalfStandHota}{1}
\newcommand{\senColHalfStandPtr}{0}
\newcommand{\senColTwoName}{colour EMA $\times 2$}
\newcommand{\senColTwoValue}{$0.3\rightarrow 0.6$}
\newcommand{\senColTwoMeanHota}{$+0.0011$}
\newcommand{\senColTwoMeanPtr}{$+0.0020$}
\newcommand{\senColTwoMeanAbsHota}{0.0019}
\newcommand{\senColTwoMaxHota}{0.0271}
\newcommand{\senColTwoMaxHotaCell}{RES\_B\_4K s8}
\newcommand{\senColTwoMaxPtr}{0.0331}
\newcommand{\senColTwoMaxPtrCell}{RES\_B\_4K s8}
\newcommand{\senColTwoStandHota}{0}
\newcommand{\senColTwoStandPtr}{1}
\newcommand{\senMaxHota}{0.0029}
\newcommand{\senMaxPtr}{0.0045}
\newcommand{\senMostSensitive}{colour EMA}
\newcommand{\senMaxHotaNoColour}{0.0019}
\newcommand{\senMaxPtrNoColour}{0.0023}
\newcommand{\senColHalfRbfkHota}{$-0.0408$}
\newcommand{\senColHalfRbfkPtr}{$-0.0713$}
\newcommand{\senRbfkMaxHota}{0.0408}
\newcommand{\senRbfkBaseHota}{0.610}
\newcommand{\senRbfkBasePtr}{0.624}
\newcommand{\senRbfkMainHota}{0.741}
\newcommand{\senRbfkMainPtr}{0.785}
\newcommand{\senMaxHotaNoRbfk}{0.0138}
\newcommand{\senMaxPtrNoRbfk}{0.0227}
\newcommand{\scrShipMean}{0.7365}
\newcommand{\scrBFail}{1}
\newcommand{\scrBMean}{0.7530}
\newcommand{\scrBtFail}{0}
\newcommand{\scrBtMean}{0.7703}
\newcommand{\scrCtFail}{3}
\newcommand{\scrCtMean}{0.7710}
\newcommand{\scrCFail}{6}
\newcommand{\scrCMean}{0.7532}
\newcommand{\btnrNonFloorFail}{14}
\newcommand{\btnrMean}{0.7675}
\newcommand{\wsfQOneNFail}{6}
\newcommand{\wsfQOneDropMin}{0.0025}
\newcommand{\wsfQOneDropMax}{0.0049}
\newcommand{\wsfQOneNFloor}{4}
\newcommand{\wsfQTwoNCells}{6}
\newcommand{\wsfQTwoNMeet}{0}
\newcommand{\wsfQTwoSwapsCells}{0}
\newcommand{\wsfQTwoSwapsSynth}{1}
\newcommand{\confKThreeHotaOff}{0.9914}
\newcommand{\confKThreeHotaOn}{0.9911}
\newcommand{\confKThreeLatOff}{13.6}
\newcommand{\confKThreeLatOn}{16.8}
\newcommand{\predRhoEd}{+0.72}
\newcommand{\predRhoMargin}{-0.59}
\newcommand{\predRhoTgran}{-0.76}
\newcommand{\predPTgran}{1.1\times10^{-6}}
\newcommand{\predEAgitMin}{0.001}
\newcommand{\predEAgitMax}{0.02}
\newcommand{\predESmoothMin}{0.11}
\newcommand{\predESmoothMax}{0.13}
\newcommand{\reproLinux}{With the pinned environment (\texttt{requirements-lock.txt}), CPU and GPU give byte-identical tracks on all $\nCells$ benchmark cells on Windows, and Linux CPU matches Windows byte for byte on the 10 cells checked.}
\newcommand{\lrnRxNcMean}{0.533}
\newcommand{\collRxNcHotaOne}{0.999}
\newcommand{\collRxNcPtrOne}{1.000}
\newcommand{\collRxNcHotaFour}{0.966}
\newcommand{\collRxNcPtrFour}{0.990}
\newcommand{\collRxNcHotaEight}{0.604}
\newcommand{\collRxNcPtrEight}{0.613}
\newcommand{\lrnRxColN}{59}
\newcommand{\lrnRxColMean}{0.565}
\newcommand{\collRxColHotaOne}{0.999}
\newcommand{\collRxColPtrOne}{1.000}
\newcommand{\collRxColHotaFour}{0.978}
\newcommand{\collRxColPtrFour}{0.991}
\newcommand{\collRxColHotaEight}{0.639}
\newcommand{\collRxColPtrEight}{0.635}
\newcommand{\lrnRxColWMean}{0.564}
\newcommand{\agRxPtrMin}{0.985}
\newcommand{\agRxPtrMax}{0.989}
\newcommand{\crowdKThreeRxMargin}{-0.001}
\newcommand{\maxGapAgCrowd}{0.003}
\newcommand{\rigN}{63}
\newcommand{\rigRho}{+0.56}
\newcommand{\rigP}{9.1\times10^{-7}}
\newcommand{\rigRawRho}{+0.74}
\newcommand{\rigRdRho}{-0.81}
\newcommand{\rigTercileIn}{4}
\newcommand{\rigTercileOf}{7}
\newcommand{\rigRMedian}{0.996}
\newcommand{\rigNHigh}{45}
\newcommand{\rigMNeg}{33}
\newcommand{\rigMPos}{24}
\newcommand{\rigMZero}{6}
\newcommand{\rigVElevenN}{48}
\newcommand{\rigVElevenRho}{+0.20}
\newcommand{\rigVElevenP}{8.5\times10^{-2}}
\newcommand{\ciBlock}{10}
\newcommand{\ciB}{1000}
\newcommand{\ciSeed}{20261005}
\newcommand{\ciLevel}{95}
\newcommand{\ciHotaAhead}{14}
\newcommand{\ciHotaAheadDrum}{2}
\newcommand{\ciHotaAheadColl}{7}
\newcommand{\ciHotaAheadAg}{2}
\newcommand{\ciHotaAheadCrowd}{0}
\newcommand{\ciHotaAheadBeds}{0}
\newcommand{\ciHotaAheadSmooth}{2}
\newcommand{\ciHotaAheadExt}{1}
\newcommand{\ciHotaBehind}{20}
\newcommand{\ciHotaBehindDrum}{2}
\newcommand{\ciHotaBehindColl}{0}
\newcommand{\ciHotaBehindAg}{3}
\newcommand{\ciHotaBehindCrowd}{0}
\newcommand{\ciHotaBehindBeds}{1}
\newcommand{\ciHotaBehindSmooth}{6}
\newcommand{\ciHotaBehindExt}{8}
\newcommand{\ciHotaNS}{25}
\newcommand{\ciHotaNSDrum}{2}
\newcommand{\ciHotaNSColl}{0}
\newcommand{\ciHotaNSAg}{0}
\newcommand{\ciHotaNSCrowd}{2}
\newcommand{\ciHotaNSBeds}{4}
\newcommand{\ciHotaNSSmooth}{15}
\newcommand{\ciHotaNSExt}{2}
\newcommand{\ciPtrAhead}{10}
\newcommand{\ciPtrAheadDrum}{3}
\newcommand{\ciPtrAheadColl}{1}
\newcommand{\ciPtrAheadAg}{1}
\newcommand{\ciPtrAheadCrowd}{0}
\newcommand{\ciPtrAheadBeds}{0}
\newcommand{\ciPtrAheadSmooth}{4}
\newcommand{\ciPtrAheadExt}{1}
\newcommand{\ciPtrBehind}{31}
\newcommand{\ciPtrBehindDrum}{1}
\newcommand{\ciPtrBehindColl}{0}
\newcommand{\ciPtrBehindAg}{4}
\newcommand{\ciPtrBehindCrowd}{1}
\newcommand{\ciPtrBehindBeds}{3}
\newcommand{\ciPtrBehindSmooth}{13}
\newcommand{\ciPtrBehindExt}{9}
\newcommand{\ciPtrNS}{18}
\newcommand{\ciPtrNSDrum}{2}
\newcommand{\ciPtrNSColl}{6}
\newcommand{\ciPtrNSAg}{0}
\newcommand{\ciPtrNSCrowd}{1}
\newcommand{\ciPtrNSBeds}{2}
\newcommand{\ciPtrNSSmooth}{6}
\newcommand{\ciPtrNSExt}{1}
\newcommand{\ciLeadSigSmooth}{2}
\newcommand{\ciCollHotaOne}{+0.0005}
\newcommand{\ciCollHotaOneLo}{+0.0002}
\newcommand{\ciCollHotaOneHi}{+0.0007}
\newcommand{\ciCollHotaFour}{+0.0030}
\newcommand{\ciCollHotaFourLo}{+0.0009}
\newcommand{\ciCollHotaFourHi}{+0.0036}
\newcommand{\ciCollHotaEight}{+0.1368}
\newcommand{\ciCollHotaEightLo}{+0.0057}
\newcommand{\ciCollHotaEightHi}{+0.2354}
\newcommand{\ciAgChuteBHota}{-0.0030}
\newcommand{\ciAgChuteBHotaLo}{-0.0049}
\newcommand{\ciAgChuteBHotaHi}{-0.0019}
\newcommand{\ciAgChuteCHota}{-0.0016}
\newcommand{\ciAgChuteCHotaLo}{-0.0028}
\newcommand{\ciAgChuteCHotaHi}{-0.0009}
\newcommand{\ciAgHopperHota}{+0.0031}
\newcommand{\ciAgHopperPtr}{-0.0132}
\newcommand{\ciAgChuteAHota}{-0.0014}
\newcommand{\ciAgChuteAHotaLo}{-0.0022}
\newcommand{\ciAgChuteAHotaHi}{-0.0008}
\newcommand{\ciLargeAHota}{-0.0158}
\newcommand{\ciLargeAHotaLo}{-0.0373}
\newcommand{\ciLargeAHotaHi}{+0.0060}
\newcommand{\ciLargeBHota}{+0.0285}
\newcommand{\ciLargeBHotaLo}{-0.0016}
\newcommand{\ciLargeBHotaHi}{+0.0365}
\newcommand{\ciResAFourKFourHota}{+0.0160}
\newcommand{\ciResAFourKFourHotaLo}{-0.0075}
\newcommand{\ciResAFourKFourHotaHi}{+0.0290}
\newcommand{\ciPtrPctCollEightLo}{95.4}
\newcommand{\ciPtrPctCollEightHi}{98.7}
\newcommand{\ciBiasHotaMed}{+0.023}
\newcommand{\ciBiasHotaMax}{+0.061}
\newcommand{\ciHotaOutside}{18}
\newcommand{\ciHotaCells}{59}
\newcommand{\ciBiasPtrMedAbs}{0.0010}
\newcommand{\ciBiasPairedMedAbs}{0.0014}
\newcommand{\regThreshold}{0.5}
\newcommand{\regEasyN}{20}
\newcommand{\regEasyMaxDef}{0.025}
\newcommand{\regLeadN}{13}
\newcommand{\regLeadPhiMin}{0.25}
\newcommand{\regLeadPhiMax}{1.7}
\newcommand{\regLeadGainMax}{0.14}
\newcommand{\regChuteFailPhiMin}{2.3}
\newcommand{\regChuteFailMaxDef}{0.25}
\newcommand{\ptrTie}{0.005}
\newcommand{\ptrClip}{0.15}
\newcommand{\ptrRegEasyN}{20}
\newcommand{\ptrRegEasyMaxDef}{0.034}
\newcommand{\ptrRegLeadPhiMin}{0.57}
\newcommand{\ptrRegLeadNonFloorN}{11}
\newcommand{\ptrRegLeadNonFloorPhiMax}{1.71}
\newcommand{\ptrRegLeadGainMax}{0.162}
\newcommand{\ptrChuteFailN}{7}
\newcommand{\ptrChuteFailPhiMin}{2.3}
\newcommand{\ptrChuteFailMaxDef}{0.70}
\newcommand{\ptrChuteFailCbtvZeroN}{4}
\newcommand{\tstatW}{1.5}
\newcommand{\tstatCasesN}{12}
\newcommand{\tstatYieldTol}{0.05}
\newcommand{\tstatCovMinPct}{98}
\newcommand{\tstatEdgeD}{20}
\newcommand{\tstatTvRatioMin}{2}
\newcommand{\tstatTvRatioMax}{234}
\newcommand{\tstatTvRatioMed}{14}
\newcommand{\tstatChuteTKEMaxPct}{6}
\newcommand{\tstatOracleTChuteMinPct}{9}
\newcommand{\tstatOracleTChuteMaxPct}{475}
\newcommand{\tstatCbtvMoreN}{3}
\newcommand{\tstatCbtvMorePhiMin}{0.38}
\newcommand{\tstatCbtvMorePhiMax}{0.97}
\newcommand{\tstatCbtvMoreGainNvpMin}{2}
\newcommand{\tstatCbtvMoreGainNvpMax}{80}
\newcommand{\tstatChuteN}{5}
\newcommand{\tstatChuteTieN}{3}
\newcommand{\tstatChutePtrGapMax}{0.030}
\newcommand{\tstatChuteRatioMin}{3}
\newcommand{\tstatChuteRatioMax}{31}
\newcommand{\tstatDrumSFourCbtvPtr}{0.71}
\newcommand{\tstatDrumSFourGothPtr}{0.30}
\newcommand{\tstatDrumSFourCbtvT}{0.98}
\newcommand{\tstatDrumSFourGothT}{0.92}
\newcommand{\tstatBeyondMinTPct}{64}
\newcommand{\tstatCbtvEdgeErrMinPct}{92}
\newcommand{\tstatCbtvEdgeErrMaxPct}{99.9}
\newcommand{\tstatEdgePtsMinPct}{37}
\newcommand{\tstatEdgePtsMaxPct}{40}
\newcommand{\tstatCbtvInteriorFloorRatioSecondMax}{1.2}
\newcommand{\tstatCbtvInteriorFloorRatioMax}{3.6}
\newcommand{\tstatCbtvInteriorFloorRatioMaxCase}{bump chute at step 4}
\newcommand{\tstatCbtvLongLinkMin}{1}
\newcommand{\tstatCbtvLongLinkMax}{29}
\newcommand{\tstatCbtvLongLinkLenMin}{1.7}
\newcommand{\tstatCbtvLongLinkLenMax}{2.3}
\newcommand{\tstatNvpLongLinkMax}{1}
\newcommand{\tstatChuteGtStepMax}{1.4}
\newcommand{\swapBandLo}{0.5}
\newcommand{\swapBandHi}{1.2}
\newcommand{\swapCasesN}{12}
\newcommand{\swapTransitionsK}{1627}
\newcommand{\swapKeptCbtvBand}{0.96}
\newcommand{\swapKeptNvpBand}{0.90}
\newcommand{\swapKeptGothBand}{0.80}
\newcommand{\swapSwapDeepBand}{0.02}
\newcommand{\swapLostDeepBand}{0.27}
\newcommand{\swapKeptRelaxBand}{0.96}
\newcommand{\swapPhiOnePctCbtv}{0.6}
\newcommand{\swapPhiOnePctNvp}{0.2}
\newcommand{\swapPhiOnePctGoth}{0.04}
\newcommand{\swapBumpLo}{2}
\newcommand{\swapBumpHi}{2.5}
\newcommand{\nCellsPrereg}{71}
\renewcommand{\reproLinux}{Linux CPU matches Windows byte for byte on the 10 cases checked.}

\title{\textbf{CBTV: Colour Bean Tracking Velocimetry for Dense Granular Flows}}

\author{
Kaveeshan Thurairajah$^{1}$\thanks{Corresponding author: \texttt{kavee@thurairajah.org}},
Zhao Pan$^{1}$\\[6pt]
$^{1}$Department of Mechanical and Mechatronics Engineering, University of Waterloo, \\
Waterloo, ON N2L 3G1, Canada \\
}

\date{}

\begin{document}

\maketitle
\begin{abstract}
Conventional particle tracking velocimetry (PTV) has been the standard diagnostic tool for studying fluid and granular flows. 
However, in dense granular systems, the average distance moved by particles between frames can approach or exceed inter-particle spacing, demanding ultra-high frame rates. 
Furthermore, neighbouring particles may move together, and the commonly used velocity-based prediction fails after energetic collisions. 
Existing methods struggle to correctly track granular particles without losing identities or swapping.
We present Colour Bean Tracking Velocimetry (CBTV), a training-free multi-pass method leveraging multicoloured particles to achieve enhanced tracking.
CBTV first matches constellations of neighbouring coloured particles as single units, guided by a cross-correlation motion prior, and only then assigns the remaining particles individually.
We compare it with \nBaselines{} baseline tracker configurations (ten methods), from classical linkers to graph-neural-network and transformer models, on \nCells{} synthetic flow--displacement cases from \nFlows{} discrete-element granular flows with exact ground truth ($\dMin$--$\dMax$ diameters/frame).
CBTV matches or exceeds every tracker tested on an impulsive collision at every displacement.
Its advantage lies in dense flows in which particles move between about half and twice their spacing per frame, where links between frames are ambiguous.
On agitated shear flow, a colour relaxation tracker, which also uses neighbour context, matches or narrowly exceeds CBTV.
Where links are unambiguous, CBTV is close to the best tracker, and beyond about two to three particle diameters per frame, the range of its motion prior, trackers that compensate for bulk motion are more accurate.
For experimentalists, this means that dense granular flows of coloured particles can be tracked at displacements per frame where trackers without neighbour context lose identities, which eases the demand for ultra-high frame rates and needs no training data.
\end{abstract}
\vspace{1em}
\vspace{1em}

\section{Introduction}
\label{sec:intro}

Granular materials are among the most widely handled substances in industry, comprising nearly 80\% of raw materials in the chemical and pharmaceutical sectors~\citep{Jaeger1996}.
Despite their ubiquity, the physics governing dense granular flows remains poorly understood compared to that of Newtonian fluids.
A major barrier to progress is the difficulty of obtaining detailed particle level trajectories inside dense, rapidly deforming granular assemblies.
X-ray tomography and magnetic resonance imaging can probe opaque three-dimensional packings~\citep{Mueth2000}, but these methods are slow, expensive, and impractical for systems requiring high temporal resolution, so most measurements use optical methods on quasi-two-dimensional geometries.

Particle Image Velocimetry (PIV), originally developed for fluid flows~\citep{Adrian2011}, has been adapted for granular systems by correlating successive images of a granular bed viewed through transparent boundaries~\citep{White2003, Lueptow2000}.
It yields Eulerian velocity fields, which are interrogation window averaged and cannot follow individual tracer particles.
Particle Tracking Velocimetry (PTV) identifies particles by local brightness and links them across frames by nearest-neighbour or probabilistic assignment~\citep{Ouellette2006, Crocker1996}.
Its performance degrades sharply when inter-frame displacements become comparable to the inter-particle spacing: \citet{Malik1993} showed that reliable nearest-neighbour linking requires a spacing-to-displacement ratio of roughly two or more, and cross-correlation with relaxation~\citep{Brevis2011}, higher-order predictive search~\citep{Cierpka2013}, and Vorono\"{i}-based methods developed for granular imagery~\citep{Capart2002} relax but do not remove this bound.
Published granular experiments keep the displacement per frame small by raising the frame rate: reported values range from $0.02$ to $0.3$ diameters per frame in vibrated monolayers~\citep{Olafsen1998}, discharging silos~\citep{Garcimartin2011}, granular gases~\citep{Trittel2024}, impact~\citep{Clark2012}, and shear cells~\citep{Utter2004} (computed from the reported speeds and frame rates where not stated).
Where the frame rate cannot keep up, automated tracking has been replaced by following coloured particles by hand~\citep{Felix2007}.
In a dense packing, mean surface separation is very small relative to particle diameter, so at inter-frame particle displacements ($\Delta x$) approaching one particle diameter ($\Delta x / d \sim 1$), per-particle linking becomes ambiguous unless the motion is predictable.

\citet{Gollin2017} assess the accuracy of PIV and PTV for granular flow measurement.
The strongest PTV methods for fluids are predictive and multi-pass: Shake-the-Box~\citep{Schanz2016} predicts each particle's position in the next frame from its track history and then finds new particles in the residual images left after the tracked ones are removed, which lets it reach high seeding densities in time-resolved data.
History-based prediction assumes trajectories that are smooth on the time scale of the frame interval; in a granular flow, collisions can reverse a particle's velocity between two frames.
Granular flows pose both difficulties at once: in the same flow some particles collide and rebound violently, so their motion cannot be extrapolated from history, while some clusters move together with almost no relative motion, which removes the velocity contrast and spatial separation that helps disambiguate trajectories in dilute systems.
Current methods often struggle to successfully link particle trajectories across frames in both situations. 

A distinct PTV family links particles by \emph{neighbour consistency}: the spring model of \citet{Okamoto1995} and the match-probability relaxation methods of \citet{Baek1996} and \citet{Ohmi2000} iteratively reinforce candidate links whose neighbours move alike, assuming that small groups of particles move nearly rigidly between frames.

Colour has been explored as a discriminant between otherwise identical particles.
\citet{Bendicks2011} used dyed tracers to reduce stereo-matching ambiguity in 3D PTV, where rainbow methods encoded \emph{depth} into wavelength for volumetric velocimetry~\citep{Xiong2017}, an approach recently reviewed by \citet{Sensors2025}.
In granular systems, coloured particles have labelled populations, for example to measure segregation or to separate species in binary mixtures~\citep{Gollin2017}, and machine learning has been applied to colour-marked rod tracking in dilute granular gases~\citep{Puzyrev2020}. 
To this point, colour remains a population or depth label rather than a tool to track individual particle identities of a single species.
\citet{Jiang2018} developed a colour PTV that tracks colour-marked particles in a pseudo-two-dimensional fluidised bed by combining Vorono\"{i} tracking with relaxation-probability tracking, verified against CFD--DEM; colour there identifies and labels the particles in each image, and the links between frames are made from positions alone.
This work is the closest to our proposed method; however, it does not use the rich information provided by colour variation in clusters.
In a dense system the arrangement of colours across a group of neighbouring particles encodes structure at a higher level than the colour of any individual particle, so matching that arrangement can resolve links that positions alone leave ambiguous.

Machine-learning-based tracking methods represent the main thrust of recent work.
Deep-learning particle trackers such as DeepTrack~\citep{Midtvedt2021} address detection and localisation.
Graph neural networks have set the state of the art in fluid PTV: GotFlow3D~\citep{Liang2023} learns 3D flow motion from consecutive particle sets via recurrent graph optimal transport, and GOTrack~\citep{Wang2026} builds a complete PTV framework around it (CNN detector, GNN displacement predictor, optimal-transport linker).
These models assume and are trained on tracer particles following incompressible fluid flow, and their transfer to dense granular flows exhibiting correlated, intermittent, and frictional motion has not previously been evaluated.
For granular flow specifically, HU-FlowNet~\citep{Xu2025} links particles in a dense drum. The learned optical-flow prior provides only the context of local mean flow, and primary linking uses a Hungarian algorithm.
As generalised experimental tools, learned trackers have a further limitation: when trained on simulated flows, they implicitly assume the correctness of the very models that granular experiments are meant to test, and real training data is scarce across the wide parameter ranges those experiments span.
This motivates a tracker that needs no training.

We present Colour Bean Tracking Velocimetry (CBTV), a particle tracking method for dense granular flows at displacements approaching the inter-particle spacing, where motion is unpredictable, crowded, or both. 
CBTV uses particles of at least three distinct colours recorded with a colour camera, and combines position and colour when matching particles between frames.
Its core benefit has two parts.
First, neighbouring particles in a dense flow move coherently, that is, neither rigidly nor independently of one another, so the arrangement of a small group of particles carries information about where each of its members is, and this larger-scale structure can be used rather than each particle alone.
Second, colour makes these group patterns distinguishable from one another, so they remain recognisable at larger deformation between frames than patterns of positions alone, which extends the range of deformation and displacement over which identities can be kept.
We call these recognisable patterns of coloured particles, present in successive frames, \emph{constellations}.

Like conventional PTV, CBTV links detected particles from frame to frame, but it tracks whole particles several pixels across, whose colour is read over their area and matched as patterns among neighbours, rather than point-like tracers located by their brightness.
Similar to the Shake-the-Box method, CBTV is multi-pass, but its first pass is different.
It matches whole colour \emph{constellations} as single units, and only the remaining particles require per-particle assignment (Section~\ref{sec:method}).
Committing whole constellations first settles most of the particles in one step and leaves far fewer candidates for the per-particle pass, which is where identities are otherwise lost.
Its motion prior is measured from the images by cross-correlation (Section~\ref{sec:method:piv}), and a term built on the \emph{spatial pattern of colours} around each particle helps to assign the remaining particles.
Matching the colour pattern of a whole neighbourhood as one unit allows CBTV to keep identities where per-particle linking becomes ambiguous.
Because CBTV tracks whole coloured particles rather than tracers, its name uses \emph{bean} rather than \emph{particle}.

The present research makes three contributions:
(i) CBTV, a multi-pass, colour-enhanced tracking method for accurate particle tracking in dense granular flow, which matches colour constellations jointly before assigning the remaining particles (Section~\ref{sec:method});
(ii) a benchmark on \nFlows{} discrete-element simulation-based granular flows rendered to video with exact ground truth, on which CBTV and \nBaselines{} baseline tracker configurations (ten methods) are compared across \nCells{} flow--displacement cases covering $\dMin$--$\dMax$ diameters/frame, with detection treated as a separate step so that only the linking of particle identities from frame to frame is compared (Section~\ref{sec:validation}); it shows that CBTV's advantage lies in a band of \emph{assignment ambiguity}, measured from the ground truth as the displacement per frame relative to the spacing between neighbouring particles, and that the displacement at which learned trackers overtake classical ones depends on the flow (Section~\ref{sec:results});
and (iii) a characterisation of scaling with particle count and image resolution.

\section{Method}
\label{sec:method}

CBTV has two stages: detection of the particles and their colours in each frame, then linking of the detections into trajectories.
Linking proceeds in two passes.
A motion prior, that is, a prediction of each particle's displacement between the two frames (Section~\ref{sec:method:piv}), gives the starting point.
The first pass matches colour constellations as single units and commits most of the particles at once.
The second pass assigns the remaining particles one by one.
Colour is measured in CIELAB space throughout, because equal numerical differences in that space are close to equal perceived differences.
Section~\ref{sec:results:mech} reports what each component contributes, measured by removing it.

\subsection{Colour palette design}

Because CBTV identifies particles partly by their colour, the colour of every particle must be classified reliably from the image, despite camera noise and variations in lighting.
The palette is chosen with this in mind. For a given number of colours $N_c$ ($N_c = 3$ in this study), the colours are placed as far apart as possible in CIELAB space, so that the smallest difference between any two of them is well above the colour noise of the camera and lighting.
Three colours are a practical compromise.
They are far enough apart to be read reliably on each particle, and the number of possible colour arrangements among $n$ neighbours grows as $N_c^{\,n}$, so even a small neighbourhood carries a colour pattern that is unlikely to recur nearby.

\subsection{Particle detection}

Detection is separate from linking particle identities from frame to frame.
In every comparison of Section~\ref{sec:results}, every tracker receives the same detections, so only the linking of particle identities from frame to frame is compared.
Because a particle's colour is read from the pixels it covers, CBTV needs particles that are several pixels across rather than point-like tracers; the smallest particles in the corpus are \dpxMin{}~px in diameter (Section~\ref{sec:validation:corpus}).
For the benchmark we use Cellpose-SAM~\citep{Pachitariu2025}, a general-purpose segmentation model, without any training on our videos. Its detection $F_1$, the harmonic mean of precision and recall when detections are matched one-to-one to ground-truth particles within $5$~px ($10$~px on the doubled-resolution large chute, whose particles are twice as large), is $0.9998$ on the collision flow and ${\approx}1.00$ across the corpus.
Each particle's colour is read from the CIELAB values of the pixels inside its detected mask.
Because detection is a separate step and every tracker receives the same detections, the choice of detector does not change the order of the trackers.
Two lighter detectors, a Hough transform with a Bayesian colour classifier and a multi-channel watershed, are available; they detect fewer particles (lower $F_1$) and give the same order of trackers.

\subsection{Colour-neighbourhood matching}
\label{sec:method:kernel}

CBTV identifies a particle by the arrangement of the particles around it, that is, by their positions relative to it and by their colours.
Two implementations were built.
The first is an image-space kernel, a three-channel CIELAB template that combines an edge and intensity map with colour-weighted neighbour blobs, matched by normalised cross-correlation.
The second, the \emph{structural} matcher, compares the detected neighbour positions and colours directly as point sets, without rendering a template.
With accurate detections the two are equally accurate, and the structural matcher has no template to build, so it is the default; the kernel is kept as a documented alternative.
Two particles of the same colour are distinguished by the different colour arrangements of their neighbours.
In the final pipeline this matcher is a secondary term in the assignment cost (weight $w_k = 0.1$, Section~\ref{sec:method:assign}).
The constellation matching of the next subsection is the primary mechanism.

\subsection{Colour constellations}
\label{sec:method:const}

The main step of linking is the joint matching of \emph{colour constellations}, small, compact groups of neighbouring particles (typically 5--12) whose colour pattern is recognisable from one frame to the next.
A neighbour graph over the detections proposes the groups.
Each group is matched to the next frame as a single unit.
Starting from the per-particle motion predictions of Section~\ref{sec:method:piv}, candidate local affine transforms are scored on how well they carry the members' positions \emph{and} colours onto detections in the next frame.
A match that clears a confidence threshold \emph{commits} all members at once and removes them from the assignment that follows.
Colour contributes most at this step.
A constellation is matched as one unit, under a local affine transform, on the colours of its members together with their relative positions.
A candidate match must therefore reproduce both the colour sequence and the spatial arrangement, which a wrong group of nearby particles is far less likely to do than to pass per-particle colour checks one at a time, and one match settles many particles at once.
Because the members are committed together to distinct detections, a committed group cannot swap members among the detections it was given.
Identities within a group are exchanged only if the whole match is wrong, which is rare.
Like the neighbour-consistency methods of Section~\ref{sec:intro}, this assumes that neighbouring particles move coherently between frames.
Because each match allows a local affine transform, a group may stretch and shear and still be matched.
The neighbour-consistency methods relax per-particle match probabilities on grey-scale positions, whereas CBTV commits colour-labelled groups as single units.
The number of groups proposed per frame grows with the particle count $N$ as $\min(2048, \max(256, \lceil 0.324\,N \rceil))$.
A fixed budget that was too small for large scenes was the largest single loss of accuracy found during development (Section~\ref{sec:results:mech}).
With the motion prior of Section~\ref{sec:method:piv}, constellations commit about $94\%$ of the particles per frame on the collision flow at moderate displacement, and $99.99\%$ of those commits are correct; on the large smooth chutes they commit $63$--$77\%$, and the rest go to the per-particle assignment.
The constellation stage involves no training.

\subsection{Displacement-gated dense motion prior}
\label{sec:method:piv}

Each track gets a prediction of where its particle will be in the next frame.
The prediction starts the constellation search and centres the search gate of the per-particle assignment, the radius within which detections are considered as candidates.
The prediction comes from one of two sources, chosen per track by its observed speed.
Slow tracks use a PIV-style cross-correlation field computed from the raw pixels of consecutive frames.
In the flows studied here, the slow tracks are the particles of the dense, slowly creeping regions, such as the bulk of a chute flow or of a dense block, and the correlation field measures their collective motion directly in the current pair of frames rather than extrapolating it from the past.
Fast tracks use a constant-velocity extrapolation of their own history, which is better than image correlation once a particle has moved far enough that the local image pattern no longer matches.
The threshold between slow and fast tracks is a displacement of about $0.5$ diameters per frame ($\Delta x/d \approx 0.5$), chosen by a parameter sweep.
Three more elaborate rules for choosing between the two sources, based on the sharpness of the correlation peak and on the agreement between a track's extrapolation and the motion of its neighbours, were tested against pre-registered criteria and did not improve on the fixed threshold.
The prior uses only the images, and no ground truth enters it.

\subsection{Residual assignment}
\label{sec:method:assign}

Particles that no constellation has committed are assigned one by one.
A cost matrix over the remaining track--detection pairs is solved by the Hungarian algorithm~\citep{Kuhn1955}.
The cost of pairing track $i$ with detection $j$ is
\begin{equation}
C_{ij} = w_d\, \hat d_{ij} + w_c\, \Delta E^{*}_{ab}(i,j) + w_k\, \bigl(1 - S_{ij}\bigr),
\label{eq:cost}
\end{equation}
with $w_d = 0.8$, $w_c = 0.1$, $w_k = 0.1$, where $\hat d_{ij}$ is the distance between the predicted position of track $i$ and detection $j$ divided by the search gate, $\Delta E^{*}_{ab}(i,j)$ is the CIE76 colour difference in CIELAB space between the colour of track $i$ and that of detection $j$, and $S_{ij}$ is the structural matcher score of Section~\ref{sec:method:kernel}.
The constellation search gate is set in particle diameters, and the remaining parameters in pixels; pixel units were found to be adequate at the image scales studied (Section~\ref{sec:results:scale}).

\subsection{Track lifecycle: coasting, exits, and colour memory}
\label{sec:method:bt}

A track that misses its particle for one or more frames must be carried across the gap without losing its identity, and three mechanisms handle this.
First, a track that has lost its particle is moved each missed frame by the motion prior of Section~\ref{sec:method:piv} instead of staying at its last position (we call this coasting), so that it stays close to the particle and re-acquires it at the next detection.
Second, a coasting track that leaves the field of view is ended, rather than left waiting at the border, where it would otherwise capture the next particle to enter.
Third, the colour term of Equation~\eqref{eq:cost} compares each detection with the colour the track stored at its last match, rather than with the current frame at the track's previous position, which is mostly background once a particle moves more than about half a radius per frame.
All three mechanisms were added after the configuration first benchmarked had been frozen.
We call the configuration with them the released configuration (BT in the tables and appendices) and the earlier one the frozen configuration (v4 ship); everywhere else, ``CBTV'' means the released configuration, and the two are compared in Sections~\ref{sec:results:ablation} and~\ref{sec:results:heldout}.
The pre-registered selection of the released configuration, a fourth mechanism that was tested and rejected, and the sensitivity of the released configuration to the parameters of these mechanisms are given in Appendices~\ref{sec:si:prereg} and~\ref{sec:si:sensitivity}.

\section{Validation methodology}
\label{sec:validation}

\subsection{Synthetic granular flow corpus with exact ground truth}
\label{sec:validation:corpus}

\begin{figure}[tbp]
  \centering
  \IfFileExists{figs/fig_cases_geometry_v11.pdf}{\includegraphics[width=\textwidth,height=0.78\textheight,keepaspectratio]{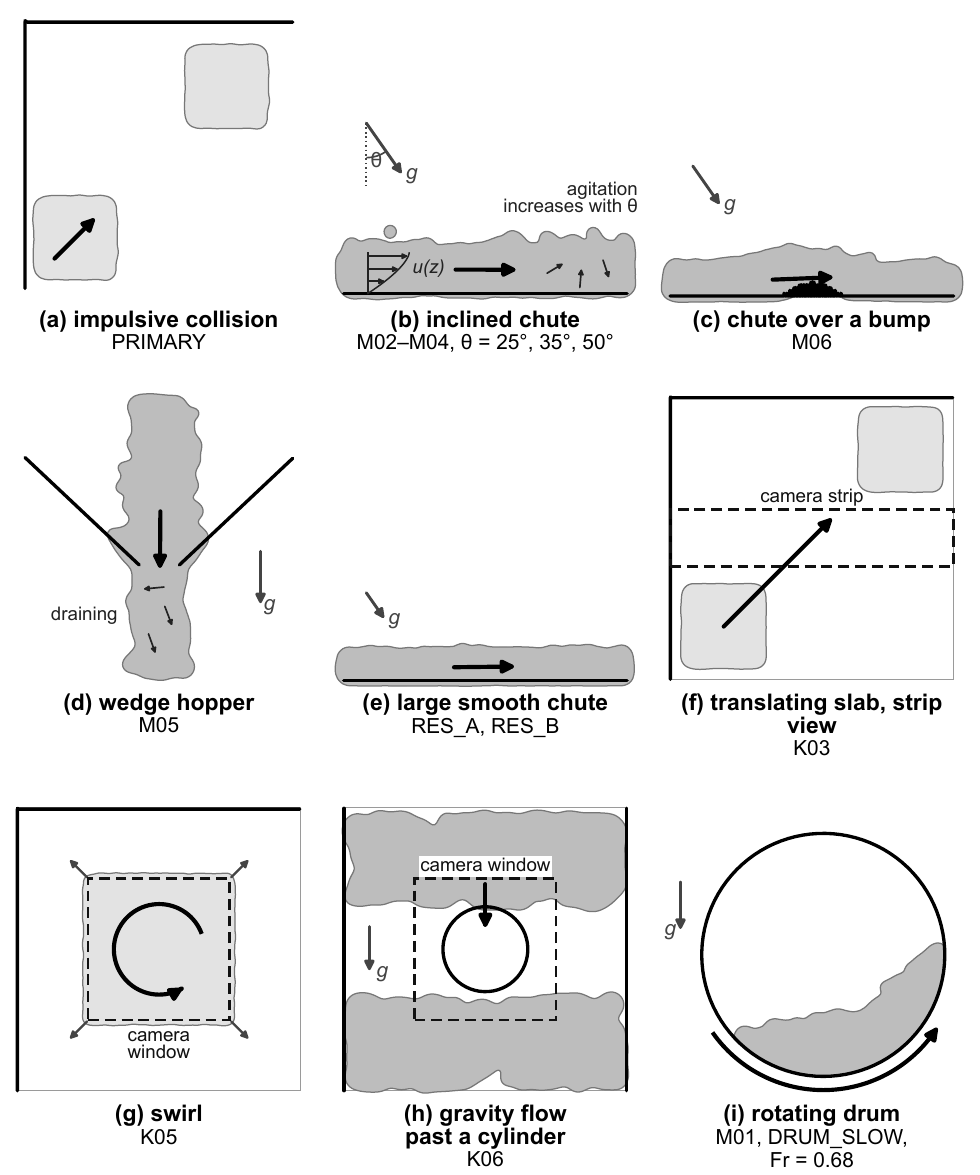}}{\framebox[\textwidth]{\rule{0pt}{3cm}\texttt{upload figs/fig\_cases\_geometry\_v11.pdf}}}
  \caption{Geometry of the flows in the corpus, traced from the wall particles and the moving particles of the simulations. Arrows show the main direction of motion and $g$ the direction of gravity. (a) Impulsive collision; (b) inclined chutes at $25^{\circ}$, $35^{\circ}$, and $50^{\circ}$; (c) chute over a bump; (d) wedge hopper; (e) large smooth chutes; (f) dense block translating through a narrow camera strip; (g) swirl; (h) gravity flow past a cylinder; (i) rotating drum. Dashed rectangles mark the camera window where it is smaller than the simulated domain. Short thin arrows in (b) and (d) indicate velocity fluctuations, and the profile in (b) is a schematic of the shear across the layer. The identifiers under each panel are those used in the tables and in Appendix~\ref{sec:si:flows}.}
  \label{fig:cases}
\end{figure}

\begin{figure}[tbp]
  \centering
  \IfFileExists{figs/fig_0_flows.pdf}{\includegraphics[width=\textwidth]{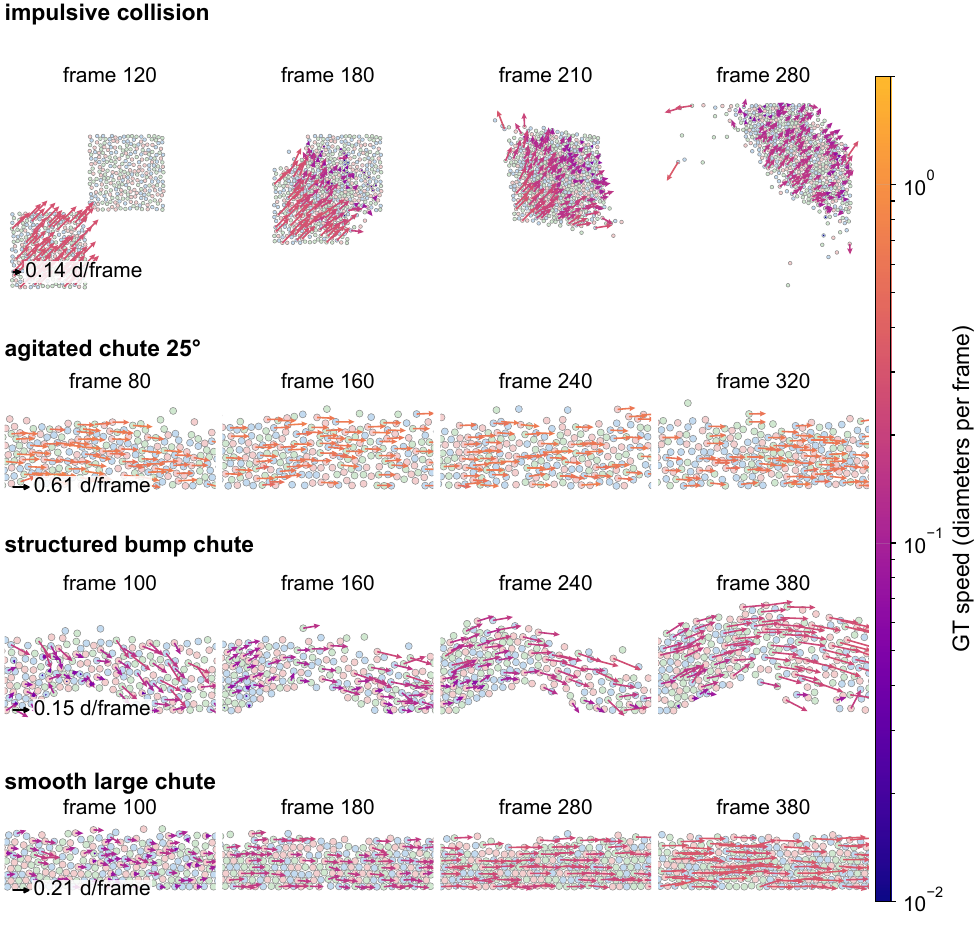}}{\framebox[\textwidth]{\rule{0pt}{3cm}\texttt{upload figs/fig\_0\_flows.pdf}}}
  \caption{The four flows of the fine displacement scan (Figure~\ref{fig:stepscan}), drawn from the ground truth at four frames each. Every particle is drawn in its colour label; arrows give the ground-truth velocity (finite difference over one frame) of a subset of particles, coloured by speed in particle diameters per frame ($d$/frame) on one logarithmic scale shared by all rows. Arrow length is scaled per row (key arrow in the first panel of each row) and is not comparable between rows. The chute rows show a 30-diameter window; every flow of the corpus is shown in Appendix~\ref{sec:si:flows}. Top row, the impulsive collision: a dense block moves into a second block at rest (frame 120), the two meet (180), merge into one compacted block (210), and eject particles (280); at contact the velocities reverse, so a constant-velocity extrapolation predicts the wrong direction.}
  \label{fig:flows}
\end{figure}

Quantitative evaluation of tracking in dense flows requires per-particle ground truth, which manual annotation cannot provide at scale.
We therefore validate on videos rendered from discrete-element simulations: the flows are simulated in LAMMPS~\citep{Thompson2022}, and the spheres are rendered to video, each with a fixed colour from the three-colour palette, on controlled backgrounds, and with controlled noise.
The corpus has fifteen flows, chosen to cover mechanically different situations: an impulsive two-body collision; sheared chutes at $25^{\circ}$/$35^{\circ}$/$50^{\circ}$; a draining wedge hopper; a chute flowing over a bump; four large-canvas smooth chutes spanning particle count ($N = 594$ to $2373$) and image scale ($\dpxLargeMin$ to $\dpxLargeMax$ px/diameter); a dense block translating through a narrow horizontal camera strip, with particles entering and leaving the window (the strip view); a coherent swirl; gravity flow past a cylinder; and a rotating drum (Froude number $0.68$), simulated at two rotation rates (two flows) with gravity scaled to hold the Froude number fixed.
The displacement per frame is controlled by subsampling the frames; its median over the particles ranges from $\dMin$ to $\dMax$ diameters per processed frame.
Each case is one flow at one sampling step, and every case is labelled by its \emph{measured} displacement, not by its nominal step.
Figure~\ref{fig:cases} sketches the geometry of every flow, and Figure~\ref{fig:flows} shows four of them in motion.

To score the trackers, the ground-truth positions from each simulation are mapped onto the pixels of its video by a fitted transform.
We check each transform by comparing every detection with its matched ground-truth position: the mean offset must be close to zero and the spread small.
We check the offsets, and not only the detection $F_1$, because a constant offset would lower every tracker's score by the same amount and would therefore not show up in the rankings.
With this check, four flows that an earlier calibration had set aside as limited by detection were found to have detections within a quarter of a pixel of the ground truth (median residual; detection $F_1$ of $1.0000$ at an 8~px match radius), and they were kept in the corpus.
The first drum flow failed the check under the earlier calibration because the fit searched too narrow a range of image scales and settled on a wrong transform.
With the range widened, it passes (detection $F_1$ of $1.0000$ at 8~px, mean offset $0.27$~px), and both drums are included.

\subsection{Metrics and protocol}

We report the particle tracking ratio (PTR) and HOTA~\citep{Luiten2021}, the standard higher-order tracking accuracy that combines detection and association.
PTR is defined over every ground-truth particle and every pair of consecutive frames as the fraction of transitions in which the particle keeps the same track identity.
A transition fails if the particle's track changes (a swap), if its track ends (lost), or if the particle has no track at all (missing), so PTR penalises lost and untracked particles as well as swaps.
It corresponds to the yield, or recovery ratio, used in PTV, the number of correct links divided by the number of ground-truth links~\citep{Baek1996}.
PTR is the main measure in the figures and in the claims made across the corpus; HOTA is given for individual cases, and the corpus-wide comparison in HOTA is in Appendix~\ref{sec:si:hota}.
Identity-swap counts are reported where they say something about the mechanism.
Four protocol rules are applied throughout.
First, \emph{detection is a separate step}.
Every tracker receives the same detections from the same detector, so any detection error affects all trackers alike, and only the linking of particle identities from frame to frame is compared.
By default, CBTV refines the centre of each detection it receives by fitting a circle to the particle's edge; setting its output positions back to the shared detections changes its HOTA and PTR by at most $0.005$ on every case, so the tables report CBTV as run.
Second, \emph{one configuration per tracker}.
Every tracker, CBTV included, was run with a single configuration on every case, with no per-video or per-step setting.
The only quantities that change between cases are rules computed from the input (CBTV's cluster budget, radius window, and motion prior, and the search radius of the relaxation tracker) and one image-scale rule that doubles the pixel gates of every tracker on the doubled-resolution chute.
The baselines were tuned on the collision flow at the smallest displacement, and the classical linkers keep the fixed search radii chosen there, $13$~px (about one particle diameter on that flow) for trackpy and LapTrack and $26$~px (about two diameters) for btrack.
CBTV's configuration was tuned on more of the corpus: its motion-prior gate on the collision flow at steps 1, 4, and 8, checked on two chute cases, and its cluster budget and weights (Section~\ref{sec:results:mech}) on three large-chute cases, which are labelled as tuning cases wherever they are quoted.
The released configuration (Section~\ref{sec:method:bt}) was selected on the corpus and confirmed on two flows simulated afterwards (Section~\ref{sec:results:heldout}).
Third, cases are ordered by measured displacement.
Fourth, deterministic trackers are run once (for a given set of detections, CBTV gives identical tracks on CPU and GPU; see Reproducibility and code availability), stochastic trackers are run at least three times, and every failure is reported.

\subsection{The baseline tracking methods}
\label{sec:validation:baselines}

CBTV is compared with nine baseline tracker configurations from four families (eight methods, one of them also in a colour-augmented variant), each run as a linker only, on the shared detections:
classical proximity and predictive PTV (trackpy nearest-neighbour linking and its nearest-velocity predictor, trackpy-NVP~\citep{Crocker1996, Allan2023}, LapTrack~\citep{Fukai2023});
Bayesian tracklet linking (btrack~\citep{Ulicna2021}, run with and without a per-particle colour feature);
video multi-object trackers (ByteTrack~\citep{Zhang2022}, OC-SORT~\citep{Cao2023}, and the learned-appearance Deep-OC-SORT~\citep{Maggiolino2023}, fed circle-to-box conversions);
and the graph-neural-network PTV linker GOTrack~\citep{Wang2026, Liang2023}, applied \emph{zero-shot}, that is, as pretrained on tracer particles in fluid flow and not fine-tuned on granular flow; where a frame held fewer detections than the 512 neighbours its correlation step expects, that number was set to the frame's detection count.
Each family relies on a different cue (proximity, velocity extrapolation, or learned appearance), so the comparison shows which cue works in which flow.
Two further learned trackers complete the set of baselines under the same protocol.
Trackastra~\citep{Gallusser2024}, a transformer linker for given detections, is applied zero-shot on all $\lrnTaN$ cases\footnote{Trackastra used its frozen window of four frames on every case except \lrnTaDevCells{} ($\lrnTaDevN$ cases, given by flow identifier and step as in Appendix~\ref{sec:si:flows}), which ran out of memory twice at that window and were run at window two, a deviation from the frozen protocol.}.
The point tracker CoTracker3~\citep{Karaev2024} must be seeded with the detections of frame 0, so it runs only on the $\lrnCoNoBirthN$ cases in which no particle enters later (Appendix~\ref{sec:si:cotracker}).
The neighbour-consistency family is represented by our own implementation of relaxation PTV from the papers~\citep{Baek1996, Ohmi2000}, since no public code exists.
It is run without colour and with the same three-bin colour label that btrack receives; colour either gates the candidate links or down-weights colour mismatches.
The colour PTV of \citet{Jiang2018} combines relaxation with Vorono\"{i} tracking and uses colour only to detect and label particles; its relaxation linker corresponds to the variant run here without colour, while the colour variant adds colour to the linking. The method itself has no public code or public validation images, so it is not compared directly.
The headline colour mode (gate) was fixed on the tuning case (collision, step 1) by a pre-registered rule before the corpus was run; the two modes tied exactly on HOTA and PTR there, and the pre-registered order decided.
Tuning on that case was already at the ceiling, so the defaults were kept for every mode.
All $\lrnRxColN$ cases completed in every mode (mean HOTA $\lrnRxNcMean$ without colour, $\lrnRxColMean$ with the gate, $\lrnRxColWMean$ with the weight).
Colour enters explicitly only for btrack and relaxation PTV, in their colour variants.
Deep-OC-SORT and CoTracker3 read the colour images and so see colour implicitly; Trackastra reads grey-level frames; the proximity, motion-only video, and graph-neural-network linkers use position alone.
The \nBaselines{} baseline tracker configurations counted throughout are these ten methods, two of them also in a colour variant; CoTracker3, which runs only on the cases in which no particle enters, is reported separately.
No single tracker is the most accurate on every flow, and an average over the corpus would depend on the mix of flows chosen; the claims below are therefore made per regime, measured as described next.

\subsection{Measured regimes of assignment ambiguity}
\label{sec:validation:regimes}

Each case is assigned to a regime by two quantities measured from the ground truth alone.
For a particle at position $\mathbf{x}_p$ in processed frame $p$, whose nearest neighbour is at distance $s_p$, the crowding is the displacement per processed frame in units of that spacing,
\begin{equation}
\phi = \frac{|\mathbf{x}_{p+1} - \mathbf{x}_p|}{s_p},
\label{eq:phi}
\end{equation}
and the unpredictability is the error of a constant-velocity prediction in the same units,
\begin{equation}
\varepsilon = \frac{|\mathbf{x}_{p+1} - (2\mathbf{x}_p - \mathbf{x}_{p-1})|}{s_p}.
\label{eq:eps}
\end{equation}
For each case we take the median and the 90th percentile of each quantity over the particles and sampled frames.
The threshold on both is $\regThreshold$, because once the true partner is displaced by more than half the nearest-neighbour spacing from the previous position or from the prediction, a wrong neighbour can be as close as the true one; this is the spacing-to-displacement ratio of about two found by \citet{Malik1993}.
A case is unpredictable if the median of $\varepsilon$ exceeds the threshold.
Otherwise, it is crowded if the median of $\phi$ exceeds it (partly unpredictable if the 90th percentile of $\varepsilon$ does), marginal if only the 90th percentile of $\phi$ does, and unambiguous if neither does.
Cases on which no tracker other than CBTV reaches a HOTA of $0.15$ are reported separately as floor cases.
Because the quantities are medians over a whole case, they do not register short impulsive events such as the few contact frames of the collision, which the errors per transition do capture (Section~\ref{sec:results:collision}).
This rule was fixed from the ground truth before the tracker scores were read for this analysis.\footnote{While the inputs were inspected, before the rule was written, the per-case table of the rigidity test (Appendix~\ref{sec:si:rigidity}), which also lists the HOTA of CBTV and of colour relaxation, was displayed once for some of the cases; no other score was seen.}
The descriptive flow groups used in the tables and figures (collision, agitated and smooth chutes, strip view, swirl and cylinder flow, drums) were assigned by flow identity after the results were known.
They are kept because they explain mechanisms, and the corpus-wide claims rest on the measured regimes.

\section{Results}
\label{sec:results}

\subsection{The crossover displacement depends on the flow}
\label{sec:results:regime}

\begin{figure}[tbp]
  \centering
  \IfFileExists{figs/fig_B_stepscan.pdf}{\includegraphics[width=\textwidth]{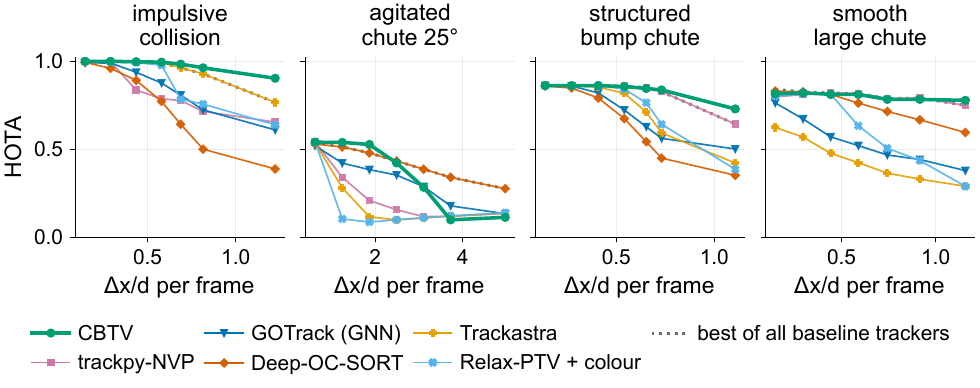}}{\framebox[\textwidth]{\rule{0pt}{3cm}\texttt{upload figs/fig\_B\_stepscan.pdf}}}
  \caption{Fine displacement scan. Each panel plots HOTA (higher-order tracking accuracy) against the measured displacement per frame, $\Delta x/d$, on a linear axis whose range differs between panels, for one flow at seven subsampling steps: the impulsive collision, the agitated $25^{\circ}$ chute, the structured bump chute and the smooth large chute. Lines: CBTV (green, bold), the five strongest other trackers (trackpy-NVP, GOTrack (GNN), Deep-OC-SORT, Trackastra, Relax-PTV with colour), and the best of all \nBaselines{} baseline trackers at each step (dotted grey); leads and ties in the text are measured against all of them. Where the crossover lies depends on the flow: on the collision CBTV ties at the first $\nCollScanTie$ steps and leads from ${\approx}\dCollFirstLead$ $d$; on the agitated chute it leads from ${\approx}\dMTwoFirstLead$ to ${\approx}\dMTwoLastLead$ $d$; on the smooth and structured chutes the extrapolator leads or ties at small steps, and CBTV leads at the largest.}
  \label{fig:stepscan}
\end{figure}

\begin{figure}[tbp]
  \centering
  \IfFileExists{figs/fig_C_ptr_vs_displacement_v12.pdf}{\includegraphics[width=\textwidth]{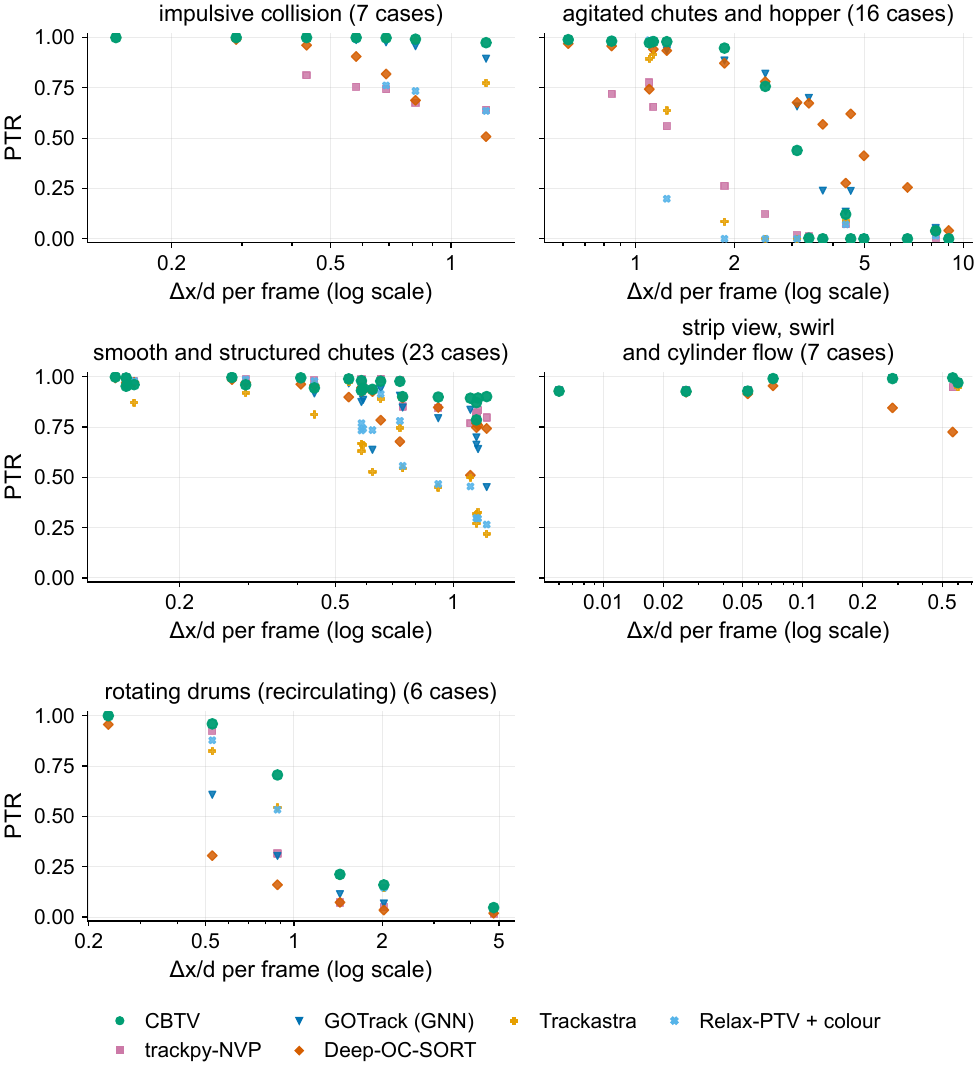}}{\framebox[\textwidth]{\rule{0pt}{3cm}\texttt{upload figs/fig\_C\_ptr\_vs\_displacement\_v11.pdf}}}
  \caption{Particle tracking ratio (PTR) against the measured median displacement per processed frame, $\Delta x/d$ (logarithmic axis; its range differs between panels), for CBTV and five other trackers (trackpy-NVP, GOTrack (GNN), Deep-OC-SORT, Trackastra, Relax-PTV with colour; one marker per tracker and case), across all \nCells{} cases of the \nFlows{}-flow corpus. One panel per flow class: the impulsive collision (7 cases), the agitated chutes and hopper (16), the smooth and structured chutes (23), the strip view, swirl and flow past a cylinder (7), and the rotating drums (6). Displacement alone does not order the outcomes, and the leader changes with the character of the flow. The same comparison in HOTA is in Figure~\ref{fig:si:hota-vs-d}.}
  \label{fig:ptr-vs-d}
\end{figure}

The displacement per frame does not by itself determine how well a tracker performs, and the crossover displacement above which learned methods overtake classical ones depends on the flow.
First, at matched ${\approx}1$ diameter/frame, outcomes differ by up to ${\sim}0.7$ HOTA depending on the flow, and the leading tracker changes (Table~\ref{tab:regime}; Figure~\ref{fig:ptr-vs-d} shows the PTR of every case).
Along the chute series the inclination increases, and with it the agitation.
The granular temperature (the fluctuating part of the velocity~\citep{Li2024}), measured from the ground truth, rises from $0.0031$ to $0.0043$ to $0.0053\,d^2$/frame$^2$.
Along this series the HOTA of the velocity-extrapolating tracker (trackpy-NVP, called the extrapolator below) falls sharply while that of CBTV changes little.
On the smooth large-canvas chutes at the same displacement the extrapolator performs well, and the two are within a few hundredths ($\regLargeABTHota$/$\regLargeATpNVPHota$ and $\regLargeBBTHota$/$\regLargeBTpNVPHota$).
Second, a fine displacement scan (seven steps per video on four flows, against all \nBaselines{} baseline trackers; Figure~\ref{fig:stepscan}) shows that where the crossover lies depends on the flow, not on the pair of trackers.
On the impulsive collision CBTV ties at the first $\nCollScanTie$ steps and leads from ${\approx}\dCollFirstLead$ $d$, with a margin that grows to $\collBTMaxScanMargin$.
On the agitated chute CBTV leads from ${\approx}\dMTwoFirstLead$ $d$ to ${\approx}\dMTwoLastLead$ $d$.
On the smooth and structured flows the extrapolator leads or ties at low displacement, and CBTV overtakes it at the largest steps (from ${\approx}\dMSixFirstLead$ $d$ on the bump chute, at ${\approx}\dResAFirstLead$ $d$ on the smooth large chute).
The agitated chute has the lowest HOTA of the four flows for every tracker, already at its smallest step (Figure~\ref{fig:stepscan}).
Identity retention there is nevertheless high (PTR $\regChuteABTPtr$ for CBTV and $\regChuteATpNVPPtr$ for trackpy-NVP, Table~\ref{tab:regime}), so the low HOTA is not caused by linking errors.

\begin{table}[H]
\centering
\caption{HOTA (PTR) at ${\approx}1$ diameter/frame for trackpy-NVP and CBTV along the chute series, in order of increasing inclination and agitation, and on the smooth large chutes. Along the chute series the HOTA of the extrapolator falls sharply while that of CBTV changes little; on the smooth large chutes the two are within a few hundredths. Last column: HOTA of CBTV minus that of the best other tracker in that case, with a paired frame-block bootstrap 95\% CI (Appendix~\ref{sec:si:ci}).}
\label{tab:regime}
\fitwidth{%
\begin{tabular}{@{}llccc@{}}
\toprule
Flow ($\Delta x/d$) & character & trackpy-NVP & CBTV & CBTV $-$ best other tracker [95\% CI] \\
\midrule
Chute $25^{\circ}$ ($\dChuteA$) & agitated & $\regChuteATpNVPHota$ ($\regChuteATpNVPPtr$) & $\regChuteABTHota$ ($\regChuteABTPtr$) & $\ciAgChuteAHota$ [$\ciAgChuteAHotaLo$, $\ciAgChuteAHotaHi$] \\
Chute $35^{\circ}$ ($\dChuteB$) & more agitated & $\regChuteBTpNVPHota$ ($\regChuteBTpNVPPtr$) & $\regChuteBBTHota$ ($\regChuteBBTPtr$) & $\ciAgChuteBHota$ [$\ciAgChuteBHotaLo$, $\ciAgChuteBHotaHi$] \\
Chute $50^{\circ}$ ($\dChuteC$) & most agitated & $\regChuteCTpNVPHota$ ($\regChuteCTpNVPPtr$) & $\regChuteCBTHota$ ($\regChuteCBTPtr$) & $\ciAgChuteCHota$ [$\ciAgChuteCHotaLo$, $\ciAgChuteCHotaHi$] \\
\midrule
Large chute ($\dLargeA$) & smooth & $\regLargeATpNVPHota$ ($\regLargeATpNVPPtr$) & $\regLargeABTHota$ ($\regLargeABTPtr$) & $\ciLargeAHota$ [$\ciLargeAHotaLo$, $\ciLargeAHotaHi$] \\
Large chute ($\dLargeB$) & smooth & $\regLargeBTpNVPHota$ ($\regLargeBTpNVPPtr$) & $\regLargeBBTHota$ ($\regLargeBBTPtr$) & $\ciLargeBHota$ [$\ciLargeBHotaLo$, $\ciLargeBHotaHi$] \\
\bottomrule
\end{tabular}}
\end{table}

The outcomes are instead organised by \emph{assignment ambiguity}, that is, by whether more than one detection is compatible with a track when it is matched, which the measured regimes of Section~\ref{sec:validation:regimes} quantify.
Ambiguity arises in two ways, from unpredictable motion and from crowding, and the corpus contains a test of each.

\subsection{Impulsive collision and agitated shear flows}
\label{sec:results:collision}

The first source of ambiguity is failure of motion prediction.
The clearest example is the impulsive two-body collision, in which a dense block of particles is driven into a second block at rest; the two meet, compact, and eject particles (Figure~\ref{fig:flows}, top row).
At impact a velocity extrapolation points the wrong way, because particles reverse and scatter, so every Kalman- and velocity-based tracker degrades.
CBTV's cross-correlation prior measures the motion after impact, and its colour constellations resolve the ejected particles.
CBTV leads or ties every other tracker at all seven displacements of the fine scan and leads from ${\approx}\dCollFirstLead$ diameters/frame (Table~\ref{tab:collision}, Figure~\ref{fig:stepscan}).
By paired 95\% confidence intervals it is ahead at all seven steps, though by less than $0.005$ HOTA at the smallest ones.
On one case (step~\errStep{}, $\errD$ diameters/frame) the lead comes from association rather than from per-frame velocity accuracy.
After impact the best other tracker's per-frame velocity errors are smaller than CBTV's, while CBTV keeps identities linked better over the whole sequence (Appendix~\ref{sec:si:velerr}).
Because the contact lasts only a few frames, the case medians of $\varepsilon$ stay below the threshold of Section~\ref{sec:validation:regimes}, and the measured regimes place the collision cases by crowding alone, from unambiguous at the smallest step to crowded at the largest.

\begin{table}[H]
\centering
\caption{Impulsive collision benchmark, detection-matched, best of each family shown. Each entry: HOTA (PTR). Trackastra is the zero-shot transformer linker; Relax-PTV is our relaxation-PTV implementation (colour: gate mode); the point tracker CoTracker3 is reported with the other cases in which no particle enters (Appendix~\ref{sec:si:cotracker}). CBTV is run once (bit-reproducible on a given platform; Appendix~\ref{sec:si:tieorder}).}
\label{tab:collision}
\fitwidth{%
\begin{tabular}{@{}lccc@{}}
\toprule
tracker & $\Delta x/d = \dCollOne$ & $\dCollFour$ & $\dCollEight$ \\
\midrule
CBTV & $\mathbf{\collBTHotaOne}$ ($\mathbf{\collBTPtrOne}$) & $\mathbf{\collBTHotaFour}$ ($\mathbf{\collBTPtrFour}$) & $\mathbf{\collBTHotaEight}$ ($\mathbf{\collBTPtrEight}$) \\
GNN & $\collGothHotaOne$ ($\collGothPtrOne$) & $\collGothHotaFour$ ($\collGothPtrFour$) & $\collGothHotaEight$ ($\collGothPtrEight$) \\
Deep-OC & $\collDeepOCHotaOne$ ($\collDeepOCPtrOne$) & $\collDeepOCHotaFour$ ($\collDeepOCPtrFour$) & $\collDeepOCHotaEight$ ($\collDeepOCPtrEight$) \\
trackpy-NVP & $\collTpNVPHotaOne$ ($\collTpNVPPtrOne$) & $\collTpNVPHotaFour$ ($\collTpNVPPtrFour$) & $\collTpNVPHotaEight$ ($\collTpNVPPtrEight$) \\
ByteTrack & $\collByteTrackHotaOne$ ($\collByteTrackPtrOne$) & $\collByteTrackHotaFour$ ($\collByteTrackPtrFour$) & $\collByteTrackHotaEight$ ($\collByteTrackPtrEight$) \\
Trackastra & $\collTaHotaOne$ ($\collTaPtrOne$) & $\collTaHotaFour$ ($\collTaPtrFour$) & $\collTaHotaEight$ ($\collTaPtrEight$) \\
Relax-PTV & $\collRxNcHotaOne$ ($\collRxNcPtrOne$) & $\collRxNcHotaFour$ ($\collRxNcPtrFour$) & $\collRxNcHotaEight$ ($\collRxNcPtrEight$) \\
Relax-PTV + colour & $\collRxColHotaOne$ ($\collRxColPtrOne$) & $\collRxColHotaFour$ ($\collRxColPtrFour$) & $\collRxColHotaEight$ ($\collRxColPtrEight$) \\
\midrule
CBTV $-$ best other tracker, HOTA [95\% CI] & $\ciCollHotaOne$ [$\ciCollHotaOneLo$, $\ciCollHotaOneHi$] & $\ciCollHotaFour$ [$\ciCollHotaFourLo$, $\ciCollHotaFourHi$] & $\ciCollHotaEight$ [$\ciCollHotaEightLo$, $\ciCollHotaEightHi$] \\
\bottomrule
\end{tabular}}
\end{table}

At $\dCollEight$ diameters/frame CBTV keeps $\collBTPtrPctEight\%$ (95\% CI $\ciPtrPctCollEightLo$--$\ciPtrPctCollEightHi\%$) of the ground-truth identities per transition, the highest of all trackers (GNN $\collGothPtrEight$, velocity extrapolation $\collTpNVPPtrEight$, learned appearance $\collDeepOCPtrEight$, Trackastra $\collTaPtrEight$).
Its HOTA margin over the best other tracker, the zero-shot transformer linker Trackastra, is $\collBTMarginFieldEight$ ($\collBTMarginGothEight$ over the GNN).
The measured regimes place the agitated shear cases (${\approx}1$ diameter/frame on the $35^{\circ}$/$50^{\circ}$ chutes and the draining hopper) with the crowded, predictable cases, since their constant-velocity prediction error is close to zero, so their ambiguity comes from crowding.
On HOTA the colour relaxation tracker is marginally ahead of CBTV on the $35^{\circ}$ and $50^{\circ}$ chutes (CBTV minus relaxation $\ciAgChuteBHota$ [$\ciAgChuteBHotaLo$, $\ciAgChuteBHotaHi$] and $\ciAgChuteCHota$ [$\ciAgChuteCHotaLo$, $\ciAgChuteCHotaHi$], significant by paired 95\% CIs).
On the hopper CBTV is marginally ahead on HOTA ($\ciAgHopperHota$) but keeps fewer identities per transition (PTR $\ciAgHopperPtr$).
Both are above every other tracker (CBTV $\agChuteBBTHota$/$\agChuteCBTHota$/$\agHopperBTHota$), and the relaxation tracker keeps more identities per transition ($\agRxPtrMin$--$\agRxPtrMax$ against CBTV's $\agBTPtrMin$--$\agBTPtrMax$; extrapolator $\agTpPtrMin$--$\agTpPtrMax$).

\subsection{Ambiguity from crowding (dense flows)}
\label{sec:results:crowding}

\begin{figure}[tbp]
  \centering
  \IfFileExists{figs/fig_regime_map_ptr_v12.pdf}{\includegraphics[width=\textwidth]{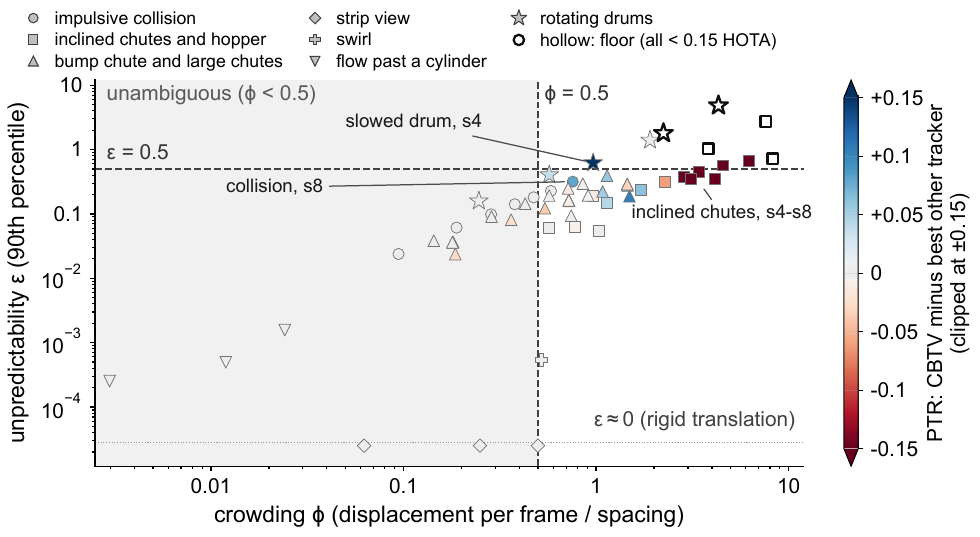}}{\framebox[\textwidth]{\rule{0pt}{3cm}\texttt{upload figs/fig\_regime\_map\_ptr\_v11.pdf}}}
  \caption{Measured regimes of the corpus. Each marker is one case, placed by the median crowding $\phi$, the displacement per processed frame in units of the nearest-neighbour spacing (horizontal axis), and by the 90th percentile of the unpredictability $\varepsilon$, the constant-velocity prediction error in the same units (vertical axis), both on logarithmic axes. Colour gives the PTR of CBTV minus that of the best other tracker in that case, clipped at ${\pm}\ptrClip$. Marker shape gives the flow group; hollow markers are floor cases, on which no tracker other than CBTV reaches a HOTA of $0.15$. The dashed lines mark the threshold of $\regThreshold$ spacings on both axes (Section~\ref{sec:validation:regimes}); cases whose prediction error is zero (rigid translation) are drawn on the dotted line at the bottom. Apart from the floor cases, CBTV's leads lie at $\phi$ between $\ptrRegLeadPhiMin$ and $\ptrRegLeadNonFloorPhiMax$.}
  \label{fig:regimes}
\end{figure}

The second source of ambiguity is crowding, that is, a displacement per frame that is a sizeable fraction of the spacing between neighbouring particles, so that a per-particle link can reach a neighbour even when the motion is predictable.
Apart from the floor cases, all $\ptrRegLeadNonFloorN$ cases on which CBTV keeps more identities than the best other tracker lie in this band, at a median $\phi$ of $\ptrRegLeadPhiMin$ to $\ptrRegLeadNonFloorPhiMax$ (Figure~\ref{fig:regimes}).
They are the collision at step 8, the large smooth chutes at steps 5 to 8, the bump chute at step 8, the $25^{\circ}$ chute at steps 2 and 3, the faster drum at step 1 and the slowed drum at step 4.
In these cases a single link is ambiguous, but a colour constellation is matched and committed as a group, so a wrong match would have to reproduce the colour arrangement of the whole group, and its members keep their identities together.
Per ground-truth transition, pooled over the $\swapCasesN$ cases with per-link outcomes, CBTV keeps $\swapKeptCbtvBand$ of the links at $\phi$ between $\swapBandLo$ and $\swapBandHi$, against $\swapKeptNvpBand$ for the extrapolator and $\swapKeptGothBand$ for the GNN (Appendix~\ref{sec:si:swaps}).
On the strip view at ${\approx}\dCrowd$ diameters/frame, where the particles move at nearly constant velocity, the velocity-extrapolating tracker makes $\swapTpKThreeEight$ identity swaps, because it estimates each velocity from its own earlier links and a wrong link carries a wrong velocity into the links that follow.
CBTV makes $\swapBTKThreeEight$ swaps there and is level with colour relaxation ($\crowdKThreeRxMargin$ HOTA), and neither of the two depends on its own velocity history.

Both sources of ambiguity lead to the same conclusion.
\textbf{Trackers that use neighbour context give the most accurate tracking, or close to it, where per-particle assignment is ambiguous, because motion is unpredictable or because the displacement approaches the inter-particle spacing, up to about $\ptrRegLeadNonFloorPhiMax$ spacings per frame. Where assignment is unambiguous, other cues perform as well or better.}
Colour relaxation's margin over CBTV tends to grow with local motion rigidity (Appendix~\ref{sec:si:rigidity}), and CBTV's constellations lead where local motion is least rigid, as in the impulsive collision at large steps.
Assignment ambiguity thus covers both the classical spacing-to-displacement criterion of \citet{Malik1993} (crowding) and flow predictability, and the cases on which joint colour-constellation assignment keeps more identities than the per-particle linkers lie within that range.
Figure~\ref{fig:regimes} places every case in the measured regimes of Section~\ref{sec:validation:regimes}, coloured by the difference in PTR; the same comparison in HOTA is in Appendix~\ref{sec:si:hota}.
Apart from the floor cases, CBTV keeps more identities than the best other tracker on $\ptrRegLeadNonFloorN$ cases, with a PTR up to $\ptrRegLeadGainMax$ higher, and all of them lie at a median displacement of $\ptrRegLeadPhiMin$ to $\ptrRegLeadNonFloorPhiMax$ nearest-neighbour spacings per frame, where links are ambiguous but the motion is still largely predictable.
On the $\ptrRegEasyN$ unambiguous and marginal cases, CBTV's PTR is at most $\ptrRegEasyMaxDef$ below that of the best other tracker.
On the $\ptrChuteFailN$ inclined-chute cases at $\ptrChuteFailPhiMin$ spacings per frame and more, the motion is close to plug flow and predictable, yet CBTV's PTR is up to $\ptrChuteFailMaxDef$ lower, and on $\ptrChuteFailCbtvZeroN$ of them it keeps no identities at all (Section~\ref{sec:results:losses}).
A difference within ${\pm}\ptrTie$ counts as a tie, and because the best other tracker is the maximum over all other tracker configurations in each case, chosen after the fact, CBTV's deficits are upper bounds.

\subsection{Where CBTV is not the best tracker}
\label{sec:results:losses}

In three regimes other methods are more accurate than CBTV; we report each with its mechanism.
\emph{Smooth flow at low displacement} is tracked best by velocity extrapolation.
On the smooth and structured chutes, CBTV trails by at most $\smBTMaxDeficit$ at low displacement (below $\smBTBehindDMax$ diameters/frame, mostly to Kalman and Bayesian linkers at the finest sampling), and from $\smBTLeadDMin$ diameters/frame it is ahead on \nBTLeadSmooth{} of the chute cases by up to $\smBTMaxLead$, significantly so on $\ciLeadSigSmooth$ (paired 95\% CIs).
The reason is that CBTV now coasts an unmatched track along the measured motion (Section~\ref{sec:method:bt}), so it carries extrapolation inside it.
The external extrapolator keeps its advantage only where every gate contains a single candidate and nothing needs to coast; it also runs at $11$--$244$~ms per frame.
At \emph{extreme displacement} (${\gtrsim}2.5$ diameters/frame) CBTV loses identities, and trackers that compensate for the bulk motion are more accurate.
Deep-OC-SORT ($\extDeepOCMin$--$\extDeepOCMax$ on the $2.5$--$5$ $d$ chute cases; beyond that it too falls to the floor value) leads every such case above the floor, while the geometric linkers are at ${\leq}\extGeoMax$ (CBTV at most $\extBTMaxThree$ above $3$ $d$), and their identity retention per transition there exceeds the geometric methods' by up to $0.7$.
At ${\approx}\dMTwoThree$ diameters/frame CBTV leads the learned-appearance tracker on the agitated chute ($\mTwoThreeBTFieldMargin$ HOTA), and at $2.5$ it trails it ($\mTwoFourBTFieldMargin$), so the boundary lies between $2$ and $2.5$ diameters/frame.
CBTV's deficit there comes from its motion prior.
The cross-correlation field measures displacements up to half its correlation window, $35$~px or about $3$ particle diameters on these chutes.
Where the bulk displacement per frame is larger, the prediction for the first frame pair falls to zero motion, no constellation commits, and each track stays at its old position and links to the particle that flows into it.
Widening the search gate does not help, because the gate stays centred on that old position.
Deep-OC-SORT keeps identities on these cases because its camera-motion compensation estimates the bulk shift of the bed and so centres its gate correctly.
GOTrack keeps identities only within its own capture range.
CBTV has a whole-frame bulk-motion pass, which is off in the released configuration.
Tested on two of these cases, it corrects the first prediction, but identities are still lost within about ten frames, and it lowers accuracy on other cases, so it stays off.
Extending the range of the motion prior remains future work.
\emph{Dense coherent rotation} is a weakness.
In the dense early phase of the swirl CBTV changes track ownership often ($2{,}114$ identity changes against the extrapolator's $29$) even though the aggregate scores stay close.
One further result is unfavourable to CBTV.
On the flow past a cylinder, which partly occludes the particles, the learned-appearance tracker scores higher at fine sampling ($\occlDeepOC$ vs $\occlBT$).

\subsection{Identity swaps and flow statistics}
\label{sec:results:stats}

\begin{figure}[tbp]
  \centering
  \IfFileExists{figs/fig_stats_cases_v12.pdf}{\includegraphics[width=\textwidth]{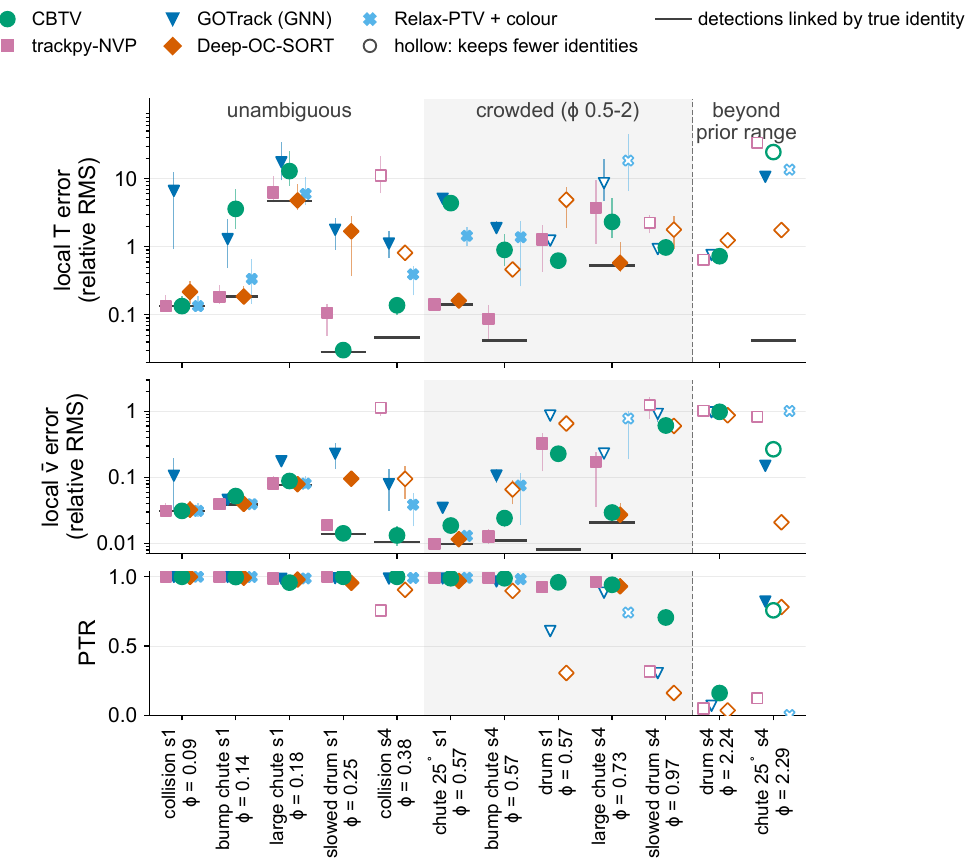}}{\framebox[\textwidth]{\rule{0pt}{3cm}\texttt{upload figs/fig\_stats\_cases\_v12.pdf}}}
  \caption{Flow statistics as an experiment would obtain them. For each of the $\tstatCasesN$ cases, ordered by crowding $\phi$ (median displacement over nearest-neighbour spacing; the shaded band is the crowded regime of Section~\ref{sec:validation:regimes}, and the cases right of the dashed line lie beyond the range of CBTV's motion prior): relative RMS error of the local granular temperature (top) and of the local mean velocity (middle), over frames and grid points holding at least three ground-truth particles within the kernel of width $\tstatW$ diameters, and the PTR of the same tracks (bottom). Marker shape gives the tracker; CBTV's positions are set back to the shared detections. Hollow markers are trackers whose PTR is more than $\tstatYieldTol$ below the highest in that case or whose coverage of the ground-truth grid is below $\tstatCovMinPct\%$, so a low error obtained by dropping particles is marked. The grey bar is the error of the shared detections linked by their true identities, the localisation floor; whiskers are 95\% block-bootstrap intervals. Relax-PTV was not run on the drums.}
  \label{fig:stats}
\end{figure}

Because identity errors bias the statistics computed from the tracks, we computed the local mean velocity, granular temperature, strain rate, and vorticity from each tracker's output on the shared detections (CBTV's positions set back to the detections, Section~\ref{sec:validation}), with a Gaussian kernel of width $\tstatW$ diameters on $\tstatCasesN$ cases, and compared each field with the same field computed from the ground truth (Figure~\ref{fig:stats}).
For the trackers that keep most identities below one spacing per frame, the local temperature error is $\tstatTvRatioMin$ to $\tstatTvRatioMax$ times the mean-velocity error of the same tracks (median $\tstatTvRatioMed$), since the temperature is a second moment of the velocity that a few wrong vectors inflate, and on the chutes at step 1, where the true temperature is below $\tstatChuteTKEMaxPct\%$ of the kinetic energy of the mean flow, even the detections linked by their true identities overstate it by $\tstatOracleTChuteMinPct$ to $\tstatOracleTChuteMaxPct\%$.
In the $\tstatCbtvMoreN$ cases in which CBTV keeps more identities than every other tracker (the collision at step 4, the drum at step 1, and the slowed drum at step 4, $\phi$ from $\tstatCbtvMorePhiMin$ to $\tstatCbtvMorePhiMax$), its temperature, strain-rate, and vorticity errors are the smallest, except for the temperature on the slowed drum at step 4, where the GNN's error is as low ($\tstatDrumSFourGothT$ against $\tstatDrumSFourCbtvT$, overlapping intervals) while it keeps $\tstatDrumSFourGothPtr$ of the identities against CBTV's $\tstatDrumSFourCbtvPtr$; against the extrapolator, CBTV's temperature error is lower by a factor of $\tstatCbtvMoreGainNvpMin$ to $\tstatCbtvMoreGainNvpMax$.
On the $\tstatChuteN$ chute cases within the range of its motion prior, where its PTR is within $\tstatChutePtrGapMax$ of the best tracker's and ties it on $\tstatChuteTieN$ of them, CBTV's temperature error is $\tstatChuteRatioMin$ to $\tstatChuteRatioMax$ times that of the best tracker (the extrapolator on the $25^{\circ}$ and bump chutes, Deep-OC-SORT on the large chute).
That excess comes from a few long links made where CBTV re-acquires a particle after coasting (Section~\ref{sec:method:bt}): $\tstatCbtvLongLinkMin$ to $\tstatCbtvLongLinkMax$ links per case of about two diameters (median $\tstatCbtvLongLinkLenMin$ to $\tstatCbtvLongLinkLenMax$ $d$, against a true step whose 99.9th percentile is at most $\tstatChuteGtStepMax$ $d$), all within $\tstatEdgeD$ diameters of the inflow or outflow, against at most $\tstatNvpLongLinkMax$ for the extrapolator.
That band holds $\tstatEdgePtsMinPct$ to $\tstatEdgePtsMaxPct\%$ of the field but $\tstatCbtvEdgeErrMinPct$ to $\tstatCbtvEdgeErrMaxPct\%$ of CBTV's squared temperature error on these cases, and in the interior its error is within a factor of $\tstatCbtvInteriorFloorRatioSecondMax$ of the detection floor on four of the five cases and $\tstatCbtvInteriorFloorRatioMax$ on the $\tstatCbtvInteriorFloorRatioMaxCase$.
Beyond two spacings per frame no tracker recovers the temperature (local errors of $\tstatBeyondMinTPct\%$ or more), and within that range CBTV improves the measured temperature, strain rate, and vorticity on the flows where it keeps more identities than the alternatives; where another tracker keeps as many, its coasting re-acquisitions at the chute ends make its temperature less accurate.

\subsection{Mechanism attribution}
\label{sec:results:mech}

Each component was ablated on the collision benchmark.
Constellation matching is the largest single contributor ($+0.19$ HOTA at $0.58$ $d$/frame over the same pipeline without it).
That ablated pipeline (a dense motion prior followed by Hungarian assignment) is structurally analogous to HU-FlowNet~\citep{Xu2025}, which links particles in a dense granular drum with an optical-flow prior and the Hungarian algorithm and uses no colour for identity; HU-FlowNet itself was not run, and its learned flow field differs from the cross-correlation field used here.
Colour is essential to the constellation matching.
A colour-blind constellation matcher drops the commit accuracy from $0.964$ to $0.681$ at $1.22$ $d$/frame.
A baseline result also indicates that colour helps as a pattern over neighbours rather than as a label per particle.
Adding a plain per-object colour feature to the Bayesian tracker destabilised its optimiser on every full-length case attempted.
Where prediction fails, the measured-motion prior has a large effect.
In an ablation on the code of that development stage, it lifts collision HOTA from $0.735$ to $0.894$ at $0.58$ $d$/frame, and replacing it by velocity extrapolation in the released configuration costs $0.04$ to $0.15$ HOTA at $0.58$ $d$ and $0.10$ to $0.24$ at $1.22$ $d$, depending on the matcher.
Across the corpus the measured prior is better or equal on 16--17 of 18 cases within the $\pm 0.005$ tie band; the exceptions are ${\leq}0.009$ on floor cases (v4 HOTA $< 0.15$; Appendix~\ref{sec:si:prereg}).
Inside CBTV, extrapolation fails through feedback.
One swap corrupts the shared prior, so the two mechanisms do not combine well in one pipeline, even though extrapolation performs well as a stand-alone tracker on smooth flow.
The largest gain during development came from the number of constellations proposed per frame.
A fixed per-frame cluster budget silently left up to a third of the particles in larger scenes to the per-particle matcher.
Raising and reweighting it recovered $0.07$--$0.15$ HOTA and improved or tied 27 of 30 original cases; scaling it with the particle count (Section~\ref{sec:method:const}) added a further $0.21$--$0.25$ on the largest scenes.
This removed what had earlier appeared to be an intrinsic decline of CBTV's advantage with particle count, which we therefore retract as an artefact of the budget.

\subsection{Scaling, resolution and compute}
\label{sec:results:scale}

With the scaled cluster budget, CBTV's accuracy holds as the particle count grows: at $N = 2373$ it matches the best other tracker on the smooth chute at moderate displacement ($\ciResAFourKFourHota$ [$\ciResAFourKFourHotaLo$, $\ciResAFourKFourHotaHi$], not significant) and scores above the zero-shot GNN on all \nLargeCells{} large-canvas cases (the four large chutes at every tested step).
Robustness to image resolution was tested from both directions.
Expressing the pixel-space parameters in particle diameters gives identical results at the scale they were tuned at and neutral or worse results elsewhere, so pixel units are not what limits CBTV at the scales studied.
On the doubled-resolution chute at step 8, CBTV scores $\resBFourKEightBT$ against $\resBFourKEightBest$ for the best other tracker (Section~\ref{sec:results:ablation}).
CBTV is slower than the baselines.
The tracking-only cost is about $0.9$--$1.3$~ms per assigned particle, against $0.02$--$0.1$ for the classical extrapolator and about $0.3$ for the GNN, and it grows with $N$.
CBTV's advantage where assignment is ambiguous is therefore in accuracy rather than in speed.

\subsection{Ablation of the released configuration}
\label{sec:results:ablation}

\begin{figure}[tbp]
  \centering
  \IfFileExists{figs/fig_E_ablation_v11.pdf}{\includegraphics[width=\textwidth]{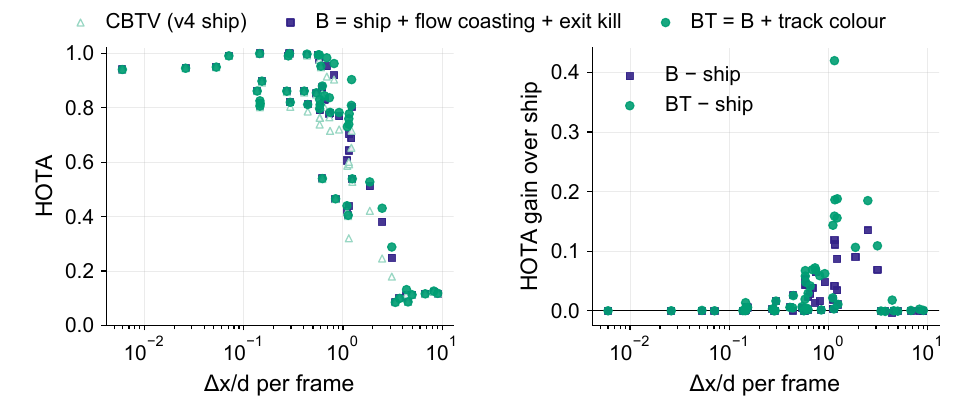}}{\framebox[\textwidth]{\rule{0pt}{3cm}\texttt{upload figs/fig\_E\_ablation.pdf}}}
  \caption{Ablation of the released configuration on the \ablNCells{} cases without the drums. Left: HOTA against the measured displacement per frame, $\Delta x/d$ (logarithmic axis), for the v4 ship configuration (hollow triangles), B = ship with flow-prior coasting and exit kill (purple squares), and BT = B with track-memory colour (green circles). Right: the HOTA gain of B and of BT over ship on the same cases. The gains concentrate at large displacement on crowded or smooth flow; BT is worse than ship by more than $0.002$ HOTA on \ablBTWorse{} cases (Table~\ref{tab:ablation}).}
  \label{fig:ablation}
\end{figure}

Table~\ref{tab:ablation} and Figure~\ref{fig:ablation} separate the two parts of the released configuration (Section~\ref{sec:method:bt}) over the \ablNCells{} cases without the drums, which were not part of the ablation.
Adding coasting and exit handling to the frozen configuration (configuration B in the table) raises mean HOTA from $\ablShipMean$ to $\ablBMean$; adding the stored track colour (BT, the released configuration) raises it to $\ablBTMean$.
BT improves \ablBTImproved{} cases by more than $0.005$ HOTA and is worse by more than $0.002$ on \ablBTWorse{} cases.
The gains concentrate at large displacement on crowded or smooth flow, for example the $25^{\circ}$ chute at step 4 ($\ablMTwoFourShip \rightarrow \ablMTwoFourB \rightarrow \ablMTwoFourBT$).
A coasting track that follows the motion prior is still near its particle when it is re-detected, and the stored colour no longer depends on whether the old position still overlaps the particle.
The released configuration does not change where the two sources of ambiguity apply.
It raises CBTV inside them (collision at $\dCollEight$ d/frame: $\collVFourHotaEight \rightarrow \collBTHotaEight$) and narrows the unambiguous regime in which extrapolation is more accurate (Section~\ref{sec:results:losses}).
The one doubled-resolution case on which the frozen configuration scored far below the best other tracker (HOTA $\resBFourKEightVFour$ vs $\resBFourKEightBest$) is largely recovered by the released configuration ($\resBFourKEightBT$).
In the ablation runs on that case (the doubled-resolution large chute at step~8; run on the frozen code, so the values differ slightly from the release-tree numbers above), coasting and exit handling raise HOTA from $\ablResBFourKEightShip$ to $\ablResBFourKEightB$ (swaps $\ablResBFourKEightShipSwap \rightarrow \ablResBFourKEightBSwap$) and track-memory colour raises it further to $\ablResBFourKEightBT$ (swaps $\ablResBFourKEightBTSwap$), the larger part of the gain.
An audit of every true track--detection pair in the cost matrices of both runs shows where the gain comes from.
Between $0.5$ and $1.5$ diameters/frame the frozen configuration's colour gate removed $\tqfVFourColGateLow\%$ (0.5--1 $d$) and $\tqfVFourColGateMid\%$ (1--1.5 $d$) of the true pairs that lay well inside the distance gate, against $\tqfBTColGateLow\%$ and $\tqfBTColGateMid\%$ for the released configuration, and the share of true pairs assigned rose from $\tqfVFourAssignedLow\%$ to $\tqfBTAssignedLow\%$ and from $\tqfVFourAssignedMid\%$ to $\tqfBTAssignedMid\%$.
This is consistent with the frozen configuration reading the current frame's colour at the track's previous position, which at $\dpxMax$ px/diameter and these displacements is mostly background, and with the ablation, in which track-memory colour carries most of the gain.
Above $1.5$ diameters/frame the gain (true pairs assigned $\tqfVFourAssignedHigh\%$ $\rightarrow$ $\tqfBTAssignedHigh\%$) is not explained by the colour gate, because the distance gate binds in both configurations ($\tqfVFourBeyondDistHigh\%$ and $\tqfBTBeyondDistHigh\%$ of true pairs lie beyond it).
We attribute it to flow-predicted coasting or to later matching stages, which the audit does not test.
The same pattern holds across the smooth chutes at step 8 (B adds $\ablSmEightBGainMin$ to $\ablSmEightBGainMax$, track colour a further $\ablSmEightTGainMin$ to $\ablSmEightTGainMax$).

\begin{table}[H]
\centering
\caption{Ablation of the released configuration over the \ablNCells{} cases without the drums (mean HOTA). The three configurations were run together on the originally frozen code, so their CBTV values can differ from the main tables (release tree) by the tie-order effects of Appendix~\ref{sec:si:tieorder}.}
\label{tab:ablation}
\fitwidth{%
\begin{tabular}{@{}lc@{}}
\toprule
configuration & mean HOTA \\
\midrule
v4 ship (as originally frozen) & $\ablShipMean$ \\
B = ship + flow-prior coasting + exit kill & $\ablBMean$ \\
BT = B + track-memory colour (release) & $\ablBTMean$ \\
\bottomrule
\end{tabular}}
\end{table}

\subsection{Out-of-sample confirmation on held-out flows}
\label{sec:results:heldout}

The released configuration was found on the evaluation corpus, so its selection is post hoc.
To test it out of sample, two new flows were simulated after the configuration was frozen and were never used in development: an oscillating-piston compaction and a silo discharging past an obstacle, each at steps 1, 4, and 8, with the corpus render settings, Cellpose detections, and the same calibration gate.
The pre-registered rule (logged before the held-out videos were rendered) required BT to be within $0.002$ of the frozen v4 configuration on both HOTA and PTR in at least five of the six cases, and to raise mean HOTA.
The rule is \hoVerdict{}, with \hoNPass{} of \hoNCells{} cases passing and a mean HOTA of $\hoBTMean$ against $\hoShipMean$ (Table~\ref{tab:heldout}).
The other trackers were run on the same detections for context; they are not part of the rule.
This pre-registered result is reported as run, on the frozen code.
Re-applying the same rule with both configurations re-run on the release tree (Appendix~\ref{sec:si:tieorder}) gives the same verdict, \hoTzVerdict{}: \hoTzNPass{} of \hoNCells{} cases pass, mean HOTA $\hoTzBTMean$ vs $\hoTzShipMean$.

\begin{table}[H]
\centering
\caption{Held-out flows (pre-registered confirmation, as run on the frozen code). HOTA (PTR). ``Best context'' is the best of the \hoNCtx{} context baselines on the same detections, not part of the rule.}
\label{tab:heldout}
\fitwidth{%
\begin{tabular}{@{}lccc@{}}
\toprule
case & CBTV (BT) & CBTV (v4) & best context \\
\midrule
piston s1 & $\hoHOneOneBTHota$ ($\hoHOneOneBTPtr$) & $\hoHOneOneVFourHota$ ($\hoHOneOneVFourPtr$) & $\hoHOneOneCtxBestHota$ (\hoHOneOneCtxBestArm) \\
piston s4 & $\hoHOneFourBTHota$ ($\hoHOneFourBTPtr$) & $\hoHOneFourVFourHota$ ($\hoHOneFourVFourPtr$) & $\hoHOneFourCtxBestHota$ (\hoHOneFourCtxBestArm) \\
piston s8 & $\hoHOneEightBTHota$ ($\hoHOneEightBTPtr$) & $\hoHOneEightVFourHota$ ($\hoHOneEightVFourPtr$) & $\hoHOneEightCtxBestHota$ (\hoHOneEightCtxBestArm) \\
silo s1 & $\hoHTwoOneBTHota$ ($\hoHTwoOneBTPtr$) & $\hoHTwoOneVFourHota$ ($\hoHTwoOneVFourPtr$) & $\hoHTwoOneCtxBestHota$ (\hoHTwoOneCtxBestArm) \\
silo s4 & $\hoHTwoFourBTHota$ ($\hoHTwoFourBTPtr$) & $\hoHTwoFourVFourHota$ ($\hoHTwoFourVFourPtr$) & $\hoHTwoFourCtxBestHota$ (\hoHTwoFourCtxBestArm) \\
silo s8 & $\hoHTwoEightBTHota$ ($\hoHTwoEightBTPtr$) & $\hoHTwoEightVFourHota$ ($\hoHTwoEightVFourPtr$) & $\hoHTwoEightCtxBestHota$ (\hoHTwoEightCtxBestArm) \\
\bottomrule
\end{tabular}}
\end{table}

\section{Planned experimental deployment}
\label{sec:experiment}

A quasi-two-dimensional vibrated cell has been constructed to deploy CBTV on real data; experimental tracking results are \emph{not} reported in this paper.
The cell confines a monolayer of painted glass beads ($d = 2$~mm, $N_c = 3$ colours) between glass plates ($100 \times 50$~mm), with corrugations along one wall rectifying oscillatory motion into directed transport.
An electromagnetic shaker (20--200~Hz, 0.5--3~mm) drives the cell; a high-speed colour camera ($1280 \times 1024$, up to 2000~fps, ${\sim}40$ px/diameter) images the face.
At the packing fractions and drive amplitudes of interest, the flow is vibration-driven, intermittent, and dense, so assignment is ambiguous from both sources of Section~\ref{sec:results}; the frame rate can be set so that $\Delta x/d$ spans about $0.3$ to $2$.
The apparatus follows the vibration-driven granular transport line of the VIP-Gran instrument~\citep{Aumaitre2018}, and the method is intended for deployment aboard ESA parabolic-flight campaigns, where Lagrangian measurements of vibration-driven transport in microgravity are not available from interrogation window averaged methods such as PIV.
Of the known risks in going from synthetic to real video, partial occlusion has been measured on a synthetic flow that stresses it (Section~\ref{sec:results:losses}).
Non-uniform illumination and frame-to-frame colour jitter are further risks that this paper does not test, and the experimental campaign will test all three directly.

\section{Discussion}
\label{sec:discussion}

Benchmarked against \nBaselines{} baseline trackers across \nFlows{} ground-truth flows, CBTV keeps the most identities on $\ptrRegLeadNonFloorN$ cases outside the floor, all at a median displacement of $\ptrRegLeadPhiMin$ to $\ptrRegLeadNonFloorPhiMax$ nearest-neighbour spacings per frame, and its PTR is within $\ptrRegEasyMaxDef$ of the best wherever links are unambiguous or marginal.
It is the most accurate on the impulsive collision. On agitated shear and on the strip view, colour relaxation matches or narrowly exceeds it.
It requires coloured particles and a colour camera, and no training.
Across the benchmark, tracking difficulty is organised by assignment ambiguity, which history-based velocity prediction does not remove.
On the strip view at $\dCrowd$ $d$, where the particles move at nearly constant velocity, the velocity-extrapolating tracker makes $\swapTpKThreeEight$ identity swaps.
The three cues we tested divide the corpus at displacements that depend on the flow.
Assignment that uses neighbour context (CBTV's constellations and, on agitated and crowded flow, colour relaxation) is the most accurate where assignment is ambiguous, up to about $\ptrRegLeadNonFloorPhiMax$ spacings per frame.
At extreme displacement, beyond the range of CBTV's motion prior, the tracker that compensates for the bulk motion of the bed (Deep-OC-SORT) is the most accurate.
Velocity extrapolation is the most accurate where assignment is unambiguous, a regime now narrowed to smooth chutes at low displacement because CBTV's coasting borrows extrapolation where it helps.
The flow statistics largely follow the identity retention.
On the $\tstatCbtvMoreN$ cases where CBTV keeps more identities than every other tracker, the granular temperature, strain rate, and vorticity computed from its tracks are the most accurate, and on the chutes, where its PTR ties the best tracker's or trails it by at most $\tstatChutePtrGapMax$, its temperature error is larger and comes from a few long links made where it re-acquires particles after coasting at the inflow and outflow (Section~\ref{sec:results:stats}).
CBTV's decisions can also be inspected, since every commit can be checked against the colour and geometry constraint that produced it, which matters in physics applications where tracker errors bias statistics without warning.

\subsection{Limitations}

The method needs optical access to individually resolved particles, which restricts it to quasi-two-dimensional or surface-layer geometries, and it currently assumes near-monodisperse spheres.
It needs particles that are coloured and large enough in the image for their colour to be measured.
Across the \nDpxFlows{} benchmark flows particles are $\dpxMin$ to $\dpxMax$ px in diameter (median $\dpxMedian$; ground-truth diameter at the calibrated image scale), and the planned experiment images them at ${\sim}40$ px/diameter; smaller particles were not tested.
It is not the best tracker on smooth chutes at low displacement, where velocity extrapolation or Kalman linking leads by up to $\smBTMaxDeficit$ HOTA; beyond the range of its motion prior (about two to three diameters per frame), where a tracker that compensates for bulk motion keeps identities; or under dense coherent rotation, its measured weak spot.
Tracking-only compute is one to two orders of magnitude above the classical extrapolator's.
All quantitative results are synthetic; real-data validation is the object of the experimental campaign in progress.

\subsection{Future directions}

Three lines follow from the benchmark.
First, choosing per particle between the measured (cross-correlation) prior and velocity extrapolation according to the local ambiguity.
Three switching rules were built and rejected against pre-registered criteria, so this choice appears to need information the tracker does not currently compute; the benchmark quantifies what a successful rule would recover.
Second, extending the range of the motion prior beyond about two to three diameters per frame, since the whole-frame bulk-motion pass corrects only the first prediction (Section~\ref{sec:results:losses}).
Third, the experimental campaign (Section~\ref{sec:experiment}), including a systematic study of performance versus colour count $N_c$ and microgravity deployment.
Two further directions are untested.
Constellation matching needs only a short descriptor per particle, so colour could be replaced by another descriptor, for example in tracking cells or crowds.
And because CBTV needs no training, its tracks of real flows could supply training data for learned trackers that do not rely on simulated flows.

\section{Conclusion}
\label{sec:conclusion}

We have presented CBTV, a training-free colour bean tracking method for dense granular flows, and benchmarked it against \nBaselines{} baseline trackers across \nFlows{} ground-truth flows and \nCells{} flow--displacement cases, with detection treated as a separate step so that only the linking of particle identities from frame to frame was compared.
The benchmark does not support the view that displacement sets tracking difficulty.
Outcomes at matched displacement differ by up to ${\sim}0.7$ HOTA, the leader changes with the character of the flow, and the crossover displacement between classical and learned methods depends on the flow.
The outcomes are organised by \emph{assignment ambiguity}, which arises from unpredictable motion or from crowding.
CBTV's joint colour-constellation assignment makes it the most accurate tracker on the impulsive collision tested and keeps it close to the best on agitated shear flow and on the strip view, where colour relaxation keeps slightly more identities per transition (CBTV at most $\maxGapAgCrowd$ HOTA behind).
Measured from the ground truth, the cases on which it keeps the most identities lie at displacements of $\ptrRegLeadPhiMin$ to $\ptrRegLeadNonFloorPhiMax$ nearest-neighbour spacings per frame, and on inclined chutes beyond the range of its motion prior, about two to three diameters per frame, it loses identities.
It leads or ties at every tested displacement on an impulsive collision, leading from ${\approx}\dCollFirstLead$ diameters/frame (HOTA $\collBTHotaOne$/$\collBTHotaFour$/$\collBTHotaEight$, with the highest identity retention of all trackers).
On agitated shear flow, only a neighbour-consistency (relaxation) tracker given the same colour labels matches or narrowly exceeds it, and CBTV is ahead of every other method.
On the strip view, where a velocity-extrapolating tracker swaps identities $\swapTpKThreeEight$ times, CBTV makes no swaps and is level with that relaxation tracker.
Colour accounts for a measured share of the accuracy, and it acts as a pattern over neighbours rather than as a label on each particle.
Its released configuration was selected by a pre-registered rule and confirmed on two flows simulated afterwards.
A vibrated-cell experiment now in progress will test the method on real, intermittently driven granular flow, where assignment is ambiguous from both sources, on the way to Lagrangian granular measurements in microgravity.
For experiments on dense granular flows, these results mean that colour constellations can keep particle identities at displacements per frame where trackers without neighbour context lose them, without training data and with less need for ultra-high frame rates.

\section*{Reproducibility and code availability}
\label{sec:repro}

CBTV will be released as open-source software under the BSD-3-Clause licence, together with the code that produced every table and figure in this paper.
All CBTV results use the release default (BT, Section~\ref{sec:method:bt}); the configuration as originally frozen is available as \texttt{--preset ship-v4}.
Both use constellation matching with the particle-count-scaled cluster budget of Section~\ref{sec:method:const}, the displacement-gated dense motion prior of Section~\ref{sec:method:piv}, and the structural matcher with the weights of Equation~\eqref{eq:cost}.
Every comparison follows the detection-matched protocol of Section~\ref{sec:validation}: the shared detection files are distributed with the release so that any linker can be re-run on exactly the inputs used here.
The release carries a regression (``golden'') test: re-running the release reproduces the published tracker output byte-for-byte on seven reference cases.
On Windows (x86-64), CPU and GPU runs of CBTV give byte-identical tracks on all $\nCells$ benchmark cases, including the drums (released configuration), and on the release's regression fixtures, with every random draw seeded from a single \texttt{--seed} option.
\reproLinux{}
Every CBTV number in this paper comes from that release tree; its tie-order sensitivity against the originally frozen runs, and the effect of the two bug fixes made before release, are reported in Appendix~\ref{sec:si:tieorder}.
Simulation inputs, rendered videos, detections, and ground truth will be archived with the release.

\section*{Acknowledgements}
We thank Joseph Owen for help with the original LAMMPS simulations, Zeyan Kherani for help with the experiments, and Disha Tandon for help with rendering.
This work is partially supported by an NSERC Discovery Grant (RGPIN-2020-04486).

\bibliographystyle{plainnat} 
\bibliography{colour_ptv_refs_v2}

\appendix

\section{Supplementary: the flows of the corpus}
\label{sec:si:flows}

\begin{figure}[H]
  \centering
  \IfFileExists{figs/fig_S_flows_all_v11.pdf}{\includegraphics[width=\textwidth,height=0.85\textheight,keepaspectratio]{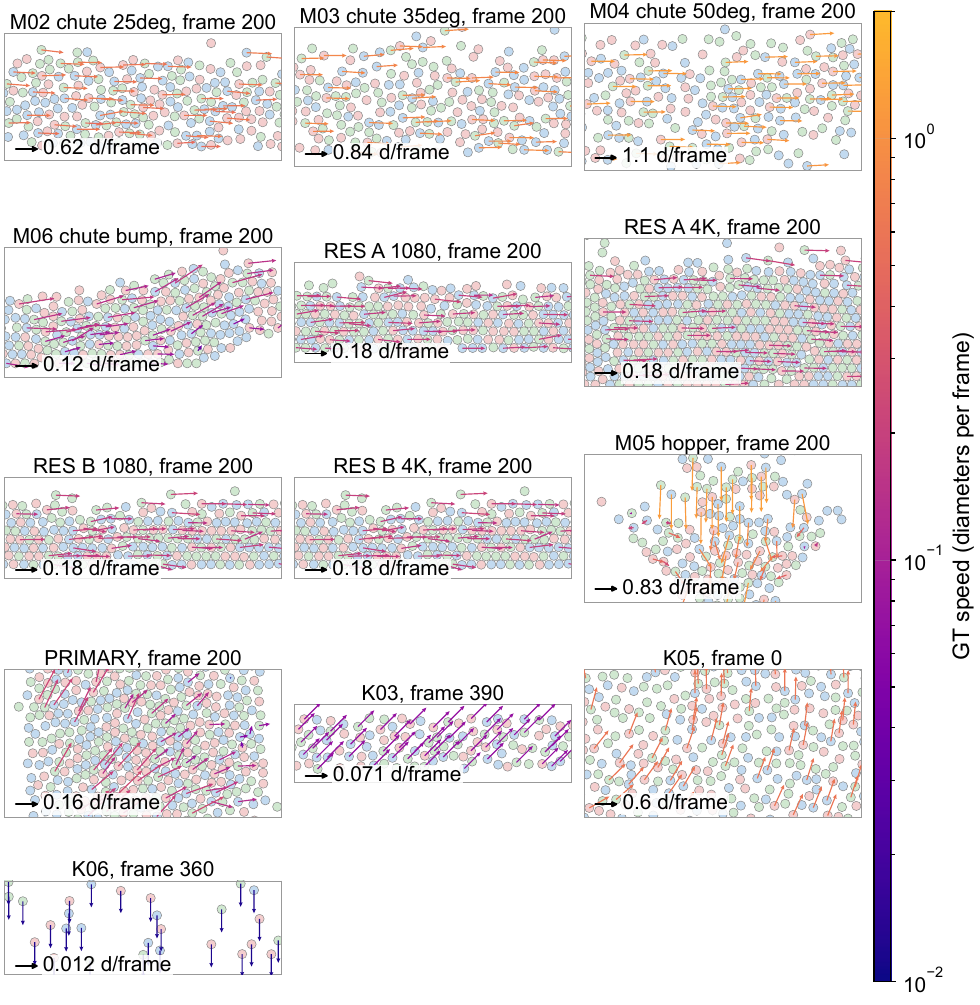}}{\framebox[\textwidth]{\rule{0pt}{3cm}\texttt{upload figs/fig\_S\_flows\_all.pdf}}}
  \caption{One ground-truth snapshot of every flow in the corpus, drawn as in Figure~\ref{fig:flows}: each panel is labelled with the flow's identifier in the data release and the frame shown, particles are drawn in their colour labels, and arrows give the ground-truth velocity of a subset of particles, coloured by speed in diameters per frame on one logarithmic scale, with a key arrow per panel.}
  \label{fig:si:flows}
\end{figure}

Figure~\ref{fig:si:flows} shows one ground-truth snapshot, drawn as in Figure~\ref{fig:flows}, of every flow in the corpus except the two rotating drums, which are sketched in Figure~\ref{fig:cases}(i).
In the supplementary tables and lists, a case is written as the flow identifier of this figure followed by its subsampling step (for example, M02 s4 is flow M02 at step 4), and RES\_B\_4K is the doubled-resolution large chute.

\onecolumn
\section{Supplementary: the corpus-wide comparison in HOTA}
\label{sec:si:hota}

\begin{figure}[H]
  \centering
  \IfFileExists{figs/fig_regime_map_hota_v12.pdf}{\includegraphics[width=\textwidth]{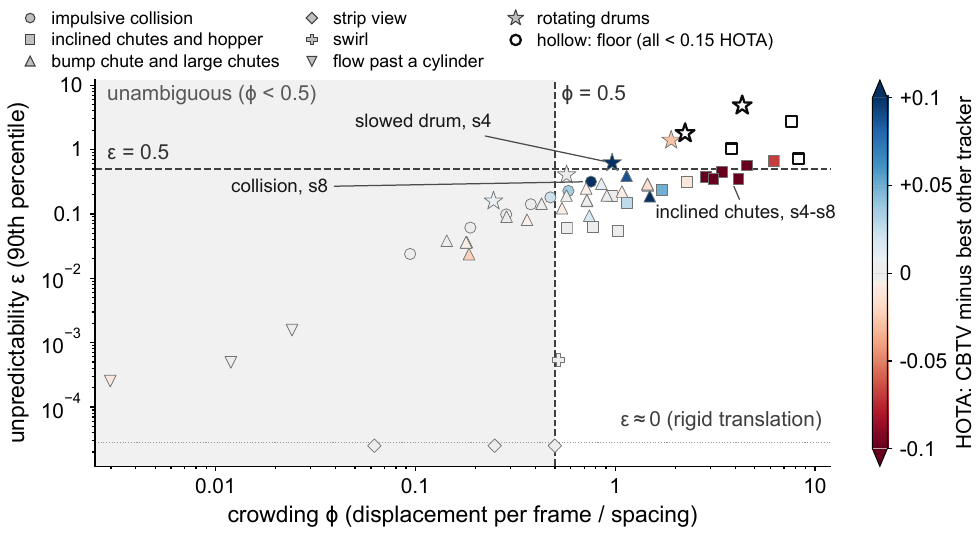}}{\framebox[\textwidth]{\rule{0pt}{3cm}\texttt{upload figs/fig\_regime\_map\_hota\_v11.pdf}}}
  \caption{Measured regimes of the corpus, drawn as in Figure~\ref{fig:regimes}, with the colour giving the HOTA of CBTV minus that of the best other tracker in each case.}
  \label{fig:si:regimes-hota}
\end{figure}

\begin{figure}[H]
  \centering
  \IfFileExists{figs/fig_C_hota_vs_displacement_v12.pdf}{\includegraphics[width=\textwidth]{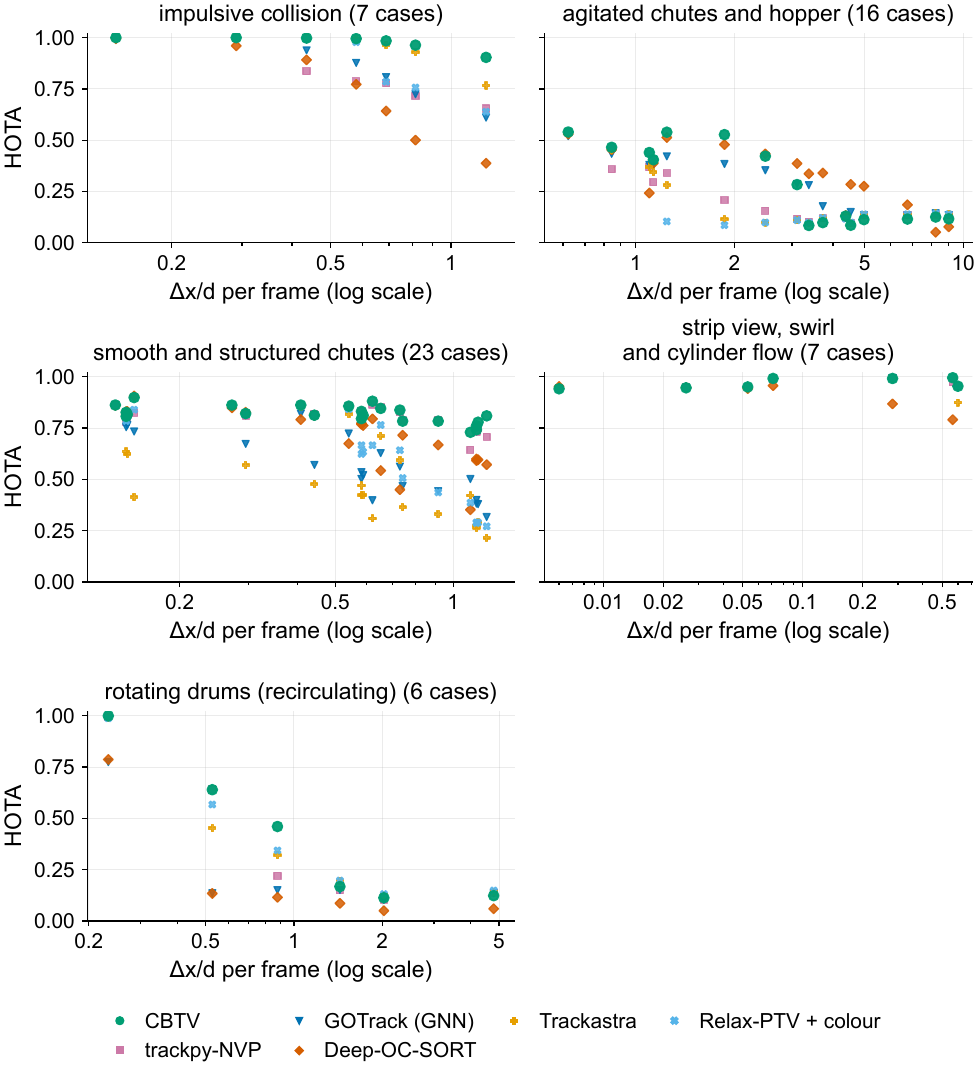}}{\framebox[\textwidth]{\rule{0pt}{3cm}\texttt{upload figs/fig\_C\_hota\_vs\_displacement\_v11.pdf}}}
  \caption{HOTA against the measured median displacement per processed frame, drawn as in Figure~\ref{fig:ptr-vs-d}.}
  \label{fig:si:hota-vs-d}
\end{figure}

Figures~\ref{fig:si:regimes-hota} and~\ref{fig:si:hota-vs-d} repeat Figures~\ref{fig:regimes} and~\ref{fig:ptr-vs-d} with HOTA in place of PTR.
In HOTA, CBTV leads the best other tracker on $\regLeadN$ cases, by up to $\regLeadGainMax$, at a median $\phi$ of $\regLeadPhiMin$ to $\regLeadPhiMax$; it is within $\regEasyMaxDef$ of the best on the $\regEasyN$ unambiguous and marginal cases, and it trails by up to $\regChuteFailMaxDef$ on the inclined chutes at $\regChuteFailPhiMin$ spacings per frame and more.

\section{Supplementary: velocity error on the collision}
\label{sec:si:velerr}

\begin{figure}[H]
  \centering
  \IfFileExists{figs/fig_err_collision.pdf}{\includegraphics[width=\textwidth]{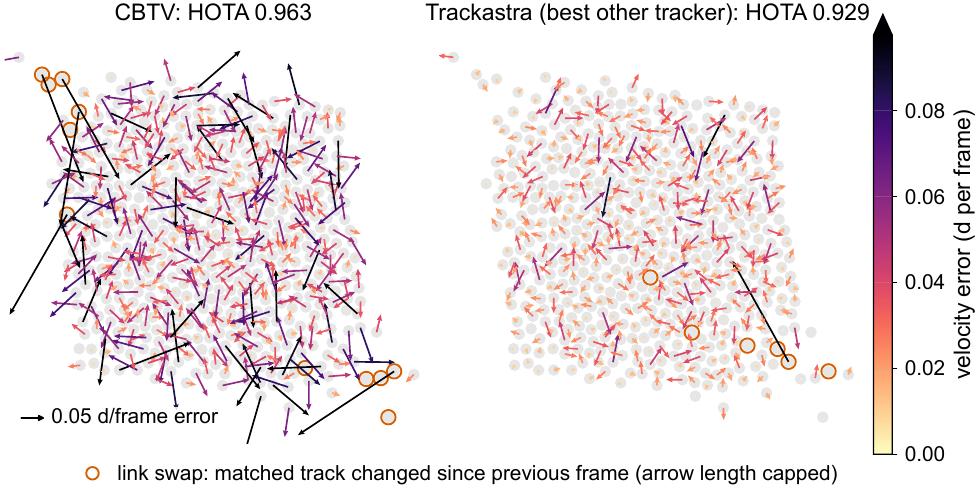}}{\framebox[\textwidth]{\rule{0pt}{3cm}\texttt{upload figs/fig\_err\_collision.pdf}}}
  \caption{Velocity error on the impulsive collision (step \errStep{}, $\errD$ diameters/frame) at processed frame \errFrame{} (source frame \errSrcFrame{}, during compression): CBTV (BT, left) and Trackastra, the best other tracker by HOTA on this case (right), with each panel's HOTA in its title. Arrows show each matched particle's velocity error (tracked minus ground-truth displacement over one processed frame), coloured by magnitude on one scale for both panels and capped in length and colour at $\errClip$ diameters/frame; grey discs are unmatched particles; rings mark post-impact link swaps, that is, particles whose matched track changed since the previous frame. Median error at this frame: $\errBTMed$ (CBTV) and $\errTaMed$ (Trackastra) diameters/frame; link swaps: $\errBTSwaps$ and $\errTaSwaps$. At this stage of the collision CBTV has the larger velocity error and more link swaps; its higher HOTA on this case comes from association over the whole sequence, as described in the text.}
  \label{fig:si:velerr}
\end{figure}

To see where CBTV's lead on the impulsive collision comes from, Figure~\ref{fig:si:velerr} compares per-particle velocity errors of CBTV (BT) and of the best other tracker on one case: the step closest to one diameter per frame (step \errStep{}, $\errD$ diameters/frame), where the best other tracker by HOTA is Trackastra ($\errTaHota$ against $\errBTHota$ for CBTV).
Tracks are matched to the ground truth as in the evaluator (within 5~px, per processed frame), and the error is the tracked minus the ground-truth displacement over one processed frame.
Before impact both trackers' errors are close to zero.
After impact CBTV is worse on both per-frame measures.
Over processed frames 30--55 its median error is larger ($\errBTMedLo$--$\errBTMedHi$ against $\errTaMedLo$--$\errTaMedHi$ diameters/frame) and it makes more \emph{post-impact link swaps} ($\errBTSwapWin$ against $\errTaSwapWin$), defined as frame transitions within frames 30--55 in which a ground-truth particle's matched track changes, under this figure's per-frame matching.
CBTV's lead on this case therefore comes from association over the whole sequence rather than from more accurate velocities.
With the same detections (detection $F_1$ $\errBTFone$ and $\errTaFone$), its association accuracy is higher (AssA $\errBTAssA$ against $\errTaAssA$) and it makes fewer \emph{evaluator identity swaps} ($\errBTSwapCell$ against $\errTaSwapCell$), the swap count of the main tables, taken over all frames of the sequence.
The two swap counts differ in window (a post-impact stretch against the whole sequence) and in matching, so they measure local and global identity errors respectively and do not contradict each other.

\section{Supplementary: where swaps begin}
\label{sec:si:swaps}

\begin{figure}[H]
  \centering
  \IfFileExists{figs/fig_swaps_vs_phi_v12.pdf}{\includegraphics[width=\textwidth]{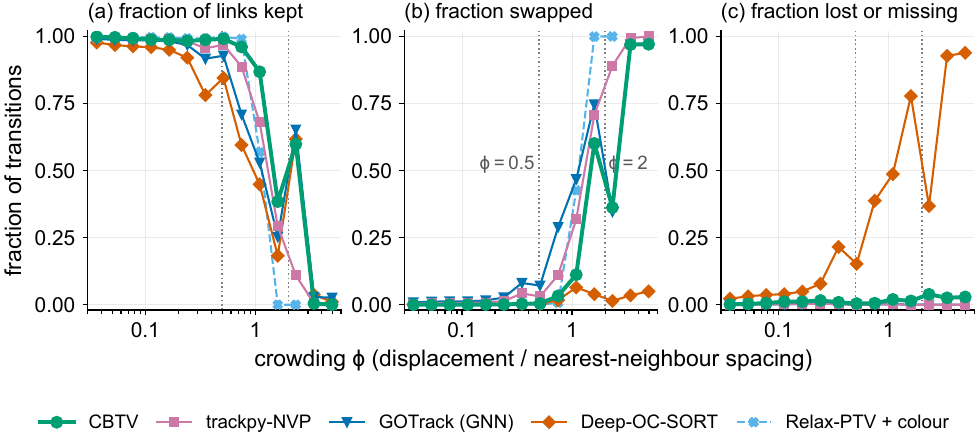}}{\framebox[\textwidth]{\rule{0pt}{3cm}\texttt{upload figs/fig\_swaps\_vs\_phi\_v12.pdf}}}
  \caption{Outcome of ground-truth transitions against crowding, pooled over the $\swapCasesN$ cases with per-link outcomes: fraction of links kept (a), swapped (b), and lost or missing (c), in bins of $\phi$ holding at least 300 transitions; (a) is the PTR per bin. The tables behind this figure keep the transitions that start inside the video frame and do not cross the periodic boundary of the chute simulations, so the chute swap counts here exclude the boundary wraps that the main-table swap counts include for every tracker alike. Relax-PTV (dashed) was not run on the drums. Dotted lines mark $\phi = 0.5$ and $2$.}
  \label{fig:si:swaps}
\end{figure}

Figure~\ref{fig:si:swaps} shows the outcome of every ground-truth particle transition of the $\swapCasesN$ cases with per-link outcomes ($\swapTransitionsK$ thousand transitions) against its crowding $\phi$, pooled over the cases.
Between $\phi = \swapBandLo$ and $\swapBandHi$, CBTV keeps $\swapKeptCbtvBand$ of the links, against $\swapKeptNvpBand$ for the extrapolator, $\swapKeptGothBand$ for the GNN, and $\swapKeptRelaxBand$ for colour relaxation on the eight non-drum cases; Deep-OC-SORT swaps only $\swapSwapDeepBand$ of the links there but loses or never tracks $\swapLostDeepBand$ of them, so its low swap rate comes with dropped links.
CBTV's swap fraction first exceeds $1\%$ at $\phi \approx \swapPhiOnePctCbtv$, the extrapolator's at $\swapPhiOnePctNvp$, and the GNN's at the smallest $\phi$ sampled ($\swapPhiOnePctGoth$).
Above $\phi \approx \swapBandHi$ every tracker loses most links; the rise of every curve between $\phi = \swapBumpLo$ and $\swapBumpHi$ is the $25^{\circ}$ chute at step 4, which dominates that bin and on which the motion is close to plug flow.

\section{Supplementary: how the released configuration was chosen}
\label{sec:si:bt}

The configuration benchmarked as ``v4 ship'' was frozen before the in/outflow work.
The three BT options were developed afterwards for particles entering and leaving the field of view; their benefit on the main corpus was a \emph{post-hoc discovery} on the evaluation corpus itself.
To limit the resulting selection bias, the choice among four candidate configurations was made by a rule written down before the candidates were scored (eligible: no case more than $0.002$ below v4 on HOTA or PTR; among eligible configurations, the highest mean HOTA; ties to the fewest flags; Appendix~\ref{sec:si:prereg}).
BT was the only eligible candidate.
It was then confirmed out of sample on two flows simulated after it was frozen (Section~\ref{sec:results:heldout}).
Two further pre-registered in/outflow questions returned negatives and are reported as such (Appendix~\ref{sec:si:prereg}: the no-harm question failed on $\wsfQOneNFail$ of $\nCellsPrereg$ cases, and the edge-benefit question was met on $\wsfQTwoNMeet$ of $\wsfQTwoNCells$ registered cases).
Both configurations are reported throughout, and the release keeps the frozen v4 configuration available as a preset that reproduces every v4 number exactly.

\section{Supplementary: a pre-registered predictability test that failed}
\label{sec:si:predictability}

Before the two sources of ambiguity of Section~\ref{sec:results} were identified, we tested a simpler hypothesis, pre-registered before any number was computed: that CBTV's advantage is explained by how predictable the motion is, measured from ground truth alone.
The predictability metric was $E_{\mathrm{extrap}}$, the median error of a constant-velocity extrapolation of the ground-truth trajectories, in particle diameters, computed on 30 cases without running any tracker; a granular-temperature proxy $T_{\mathrm{gran}}$ (a nearest-neighbour velocity-fluctuation measure) was a secondary diagnostic.
The pre-registered prediction was a positive correlation between CBTV's HOTA margin over the best other tracker and $E_{\mathrm{extrap}}$.

\begin{figure}[H]
  \centering
  \IfFileExists{figs/fig_B1_hota_vs_E_panels.pdf}{\includegraphics[width=\textwidth]{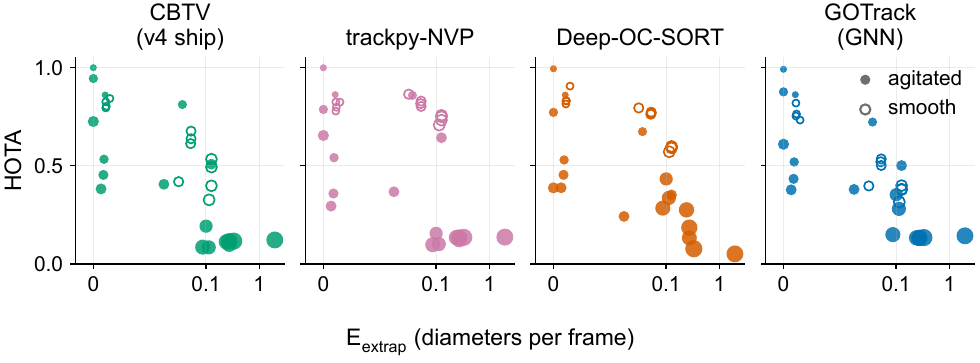}}{\framebox[\textwidth]{\rule{0pt}{3cm}\texttt{upload figs/fig\_B1\_hota\_vs\_E\_panels.pdf}}}
  \caption{HOTA against the ground-truth constant-velocity extrapolation residual $E_{\mathrm{extrap}}$ (median over particles and sampled frame triples, in diameters per processed frame; symmetric-logarithmic axis), one panel per tracker: CBTV (v4 ship), trackpy-NVP, Deep-OC-SORT, and GOTrack (GNN). Marker area grows with the measured displacement per processed frame; filled markers are the agitated cases (including the impulsive collision), open markers the smooth large chutes. $E_{\mathrm{extrap}}$ follows the sampling interval rather than the agitation (Spearman $\predRhoEd$ with $d$/frame): the agitated single-step cases lie at $E \sim \predEAgitMin$--$\predEAgitMax$ and the smooth step-8 cases at $E \sim \predESmoothMin$--$\predESmoothMax$, so a one-step extrapolation residual does not separate the regimes at matched displacement.}
  \label{fig:si:B1}
\end{figure}

\begin{figure}[H]
  \centering
  \IfFileExists{figs/fig_B2_margin_vs_E.pdf}{\includegraphics[width=0.6\textwidth]{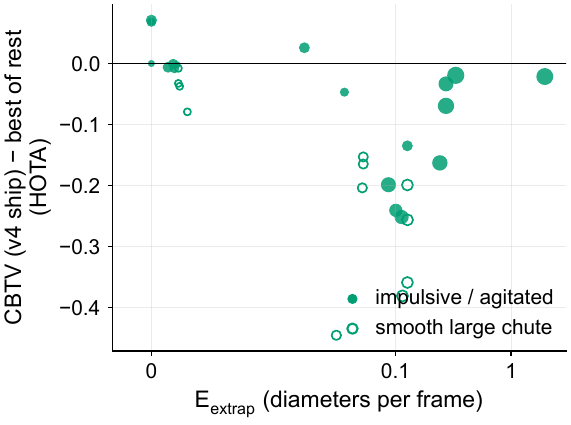}}{\framebox[0.6\textwidth]{\rule{0pt}{3cm}\texttt{upload figs/fig\_B2\_margin\_vs\_E.pdf}}}
  \caption{HOTA margin of CBTV (v4 ship, the configuration frozen at the time of the pre-registered test) over the best other tracker, per case, against $E_{\mathrm{extrap}}$ (axes as in Figure~\ref{fig:si:B1}); positive values mean that CBTV scores higher than every other tracker. Filled markers are the impulsive and agitated cases, open markers the smooth large chutes; marker area grows with displacement. The margin does not follow $E_{\mathrm{extrap}}$ (Spearman $\predRhoMargin$, the opposite sign to the prediction): CBTV's leads lie in the impulsive and agitated cases, which the interval-dominated $E_{\mathrm{extrap}}$ places at the low end of its axis.}
  \label{fig:si:B2}
\end{figure}

\begin{figure}[H]
  \centering
  \IfFileExists{figs/fig_B3_diag_hota_vs_Tgran.pdf}{\includegraphics[width=0.6\textwidth]{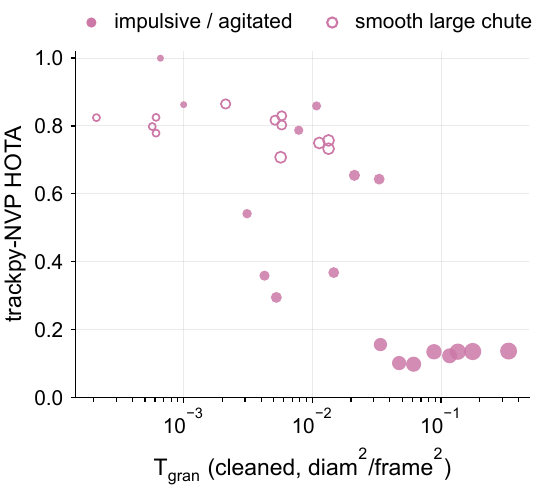}}{\framebox[0.6\textwidth]{\rule{0pt}{3cm}\texttt{upload figs/fig\_B3\_diag\_hota\_vs\_Tgran.pdf}}}
  \caption{A post-hoc diagnostic, not part of the pre-registered test: HOTA of trackpy-NVP against the granular-temperature proxy $T_{\mathrm{gran}}$ (in diameters$^2$ per frame$^2$, logarithmic axis, after removal of the periodic-box wrap from the ground-truth velocities); filled markers are the impulsive and agitated cases, open markers the smooth large chutes, and marker area grows with displacement. Spearman $r = \predRhoTgran$ ($p = \predPTgran$). The relation is a correlation rather than a separation of the regimes. trackpy-NVP predicts each particle from a neighbour's velocity, so it degrades as the velocities of neighbouring particles become less alike.}
  \label{fig:si:B3}
\end{figure}

The test failed.
The correlation was $r = -0.585$ ($p = 6.8\times10^{-4}$), the \emph{opposite sign} to the prediction.
All trackers' HOTA was dominated by displacement, and once displacement was controlled $E_{\mathrm{extrap}}$ added nothing for CBTV (partial $r = -0.145$, $p = 0.45$).
Extrapolation error grows faster than linearly with the sampling interval, so a coarsely subsampled smooth case accumulates more error than one frame of agitated flow, and the metric tracks the interval rather than the agitation.
Figures~\ref{fig:si:B1}--\ref{fig:si:B3} show the failed test and the diagnostics.

This negative result led to the explanation reported in the main text, that tracking difficulty is set by assignment ambiguity, which arises either from unpredictable motion or from crowding (Sections~\ref{sec:results:collision} and~\ref{sec:results:crowding}); a single monotone predictability score cannot capture the second source.
No claim in this paper rests on $E_{\mathrm{extrap}}$, and $T_{\mathrm{gran}}$ correlations are post hoc.

\section{Supplementary: point-tracking baseline on cases in which no particle enters}
\label{sec:si:cotracker}
CoTracker3 tracks the frame-0 detections as query points and cannot start new tracks, so it was run only on the $\lrnCoNoBirthN$ cases in which no ground-truth particle appears after the first frame (decided by script over the ground truth, never by hand): the collision, the four large-canvas chutes, the $25^{\circ}$/$35^{\circ}$/$50^{\circ}$ chutes, the hopper, the bump chute, the swirl, and the two rotating drums.
On $\lrnCoFailN$ of these cases every query point lost visibility before the end of the video, so the run produced no tracks for the remaining frames; these runs are reported as \emph{failed} and excluded from every mean.
Over the $\lrnCoN$ completed cases CoTracker3 averages HOTA $\lrnCoMean$, against $\lrnCoBTMean$ for CBTV (BT) on the same cases (Table~\ref{tab:si:cotracker}).
At the collision a sensitivity run with the model's internal resolution set to the native $1024\times1024$ instead of its default $384\times512$ raises it to $\collCoMaxresHotaOne$/$\collCoMaxresHotaFour$ (steps 1/4; $\collCoHotaOne$/$\collCoHotaFour$ at the default resolution).
Trackastra averages HOTA $\lrnTaMean$ over all $\lrnTaN$ cases, against $\lrnTaBTMean$ for CBTV (BT).

\begin{table}[H]
\centering
\caption{CoTracker3 on the cases in which no particle enters, by descriptive flow group: mean HOTA over the completed cases, with Trackastra and CBTV (BT) on the same cases. Failed runs (all query points lost before the end of the video) are counted but not averaged; their partial scores are in Table~\ref{tab:si:cotracker-failed}.}
\label{tab:si:cotracker}
\begin{tabular}{@{}lccccc@{}}
\toprule
Flow group & cases & failed & CoTracker3 & Trackastra & CBTV (BT) \\
\midrule
Impulsive collision & $\coRegCollN$ & $\coRegCollFail$ & $\coRegCollCo$ & $\coRegCollTa$ & $\coRegCollBT$ \\
Agitated chutes and hopper, $<1.5$ $d$ & $\coRegAgN$ & $\coRegAgFail$ & $\coRegAgCo$ & $\coRegAgTa$ & $\coRegAgBT$ \\
Smooth and structured chutes & $\coRegSmoothN$ & $\coRegSmoothFail$ & $\coRegSmoothCo$ & $\coRegSmoothTa$ & $\coRegSmoothBT$ \\
Swirl & $\coRegBedsN$ & $\coRegBedsFail$ & -- & -- & -- \\
Agitated chutes and hopper, $\geq 1.5$ $d$ & $\coRegExtN$ & $\coRegExtFail$ & $\coRegExtCo$ & $\coRegExtTa$ & $\coRegExtBT$ \\
Rotating drums & $\coRegDrumN$ & $\coRegDrumFail$ & $\coRegDrumCo$ & $\coRegDrumTa$ & $\coRegDrumBT$ \\
\midrule
All such cases & $\lrnCoN$ & $\lrnCoFailN$ & $\lrnCoMean$ & $\lrnCoTaMean$ & $\lrnCoBTMean$ \\
\bottomrule
\end{tabular}
\end{table}

\begin{table}[H]
\centering
\caption{The $\lrnCoFailN$ failed CoTracker3 runs. Partial run, lower bound: HOTA scored by the benchmark evaluator with only its completeness check bypassed, so every frame after the last tracked frame (index in the subsampled sequence) counts as missed. Not used in any main-text number.}
\label{tab:si:cotracker-failed}
\begin{tabular}{@{}lcccc@{}}
\toprule
Case & $\Delta x/d$ & last tracked frame & HOTA (partial, lower bound) & CBTV (BT) \\
\midrule
M03 chute $35^{\circ}$, step 1 & $\coPartMThreeOneD$ & $\coPartMThreeOneEnd$ & $\coPartMThreeOneHota$ & $\coPartMThreeOneBT$ \\
M03 chute $35^{\circ}$, step 4 & $\coPartMThreeFourD$ & $\coPartMThreeFourEnd$ & $\coPartMThreeFourHota$ & $\coPartMThreeFourBT$ \\
M03 chute $35^{\circ}$, step 8 & $\coPartMThreeEightD$ & $\coPartMThreeEightEnd$ & $\coPartMThreeEightHota$ & $\coPartMThreeEightBT$ \\
M04 chute $50^{\circ}$, step 1 & $\coPartMFourOneD$ & $\coPartMFourOneEnd$ & $\coPartMFourOneHota$ & $\coPartMFourOneBT$ \\
M04 chute $50^{\circ}$, step 4 & $\coPartMFourFourD$ & $\coPartMFourFourEnd$ & $\coPartMFourFourHota$ & $\coPartMFourFourBT$ \\
M04 chute $50^{\circ}$, step 8 & $\coPartMFourEightD$ & $\coPartMFourEightEnd$ & $\coPartMFourEightHota$ & $\coPartMFourEightBT$ \\
M05 hopper, step 8 & $\coPartMFiveEightD$ & $\coPartMFiveEightEnd$ & $\coPartMFiveEightHota$ & $\coPartMFiveEightBT$ \\
K05 swirl, step 1 & $\coPartKFiveOneD$ & $\coPartKFiveOneEnd$ & $\coPartKFiveOneHota$ & $\coPartKFiveOneBT$ \\
M01 drum, step 1 & $\coPartMOneDrumOneD$ & $\coPartMOneDrumOneEnd$ & $\coPartMOneDrumOneHota$ & $\coPartMOneDrumOneBT$ \\
Slowed drum, step 4 & $\coPartDrumSlowFourD$ & $\coPartDrumSlowFourEnd$ & $\coPartDrumSlowFourHota$ & $\coPartDrumSlowFourBT$ \\
Slowed drum, step 8 & $\coPartDrumSlowEightD$ & $\coPartDrumSlowEightEnd$ & $\coPartDrumSlowEightHota$ & $\coPartDrumSlowEightBT$ \\
\bottomrule
\end{tabular}
\end{table}

\section{Supplementary: tie-order sensitivity of the CBTV numbers}
\label{sec:si:tieorder}
The originally frozen CBTV runs used numpy's default (unstable) sort, whose tie-breaking depends on the CPU's SIMD backend, so the same code could assign different track identities (and, through cascades, occasionally different links) on different machines.
The release replaces every tie-sensitive sort with a stable sort, which makes the output identical across CPU/GPU and SIMD backends on the regression fixtures, and every CBTV number in this paper was regenerated on that tree.
Table~\ref{tab:si:tieorder} compares the stable-sort values (before the two bug fixes described below) with the originally frozen ones over the \ablNCells{} corpus cases without the drums.

\begin{table}[H]
\centering
\caption{Tie-order sensitivity: stable-sort tree (before the two release fixes) vs originally frozen runs, per case, $d = \max(|\Delta\mathrm{HOTA}|, |\Delta\mathrm{PTR}|)$.}
\label{tab:si:tieorder}
\begin{tabular}{@{}lccccc@{}}
\toprule
configuration & exact & $d \leq 0.002$ & $d > 0.002$ & max $|\Delta\mathrm{HOTA}|$ & max $d$ \\
\midrule
v4 & $\tieShipExact$ & $\tieShipWithin$ & $\tieShipBeyond$ & $\tieShipMaxHota$ & $\tieShipMaxAny$ \\
BT (released) & $\tieBTExact$ & $\tieBTWithin$ & $\tieBTBeyond$ & $\tieBTMaxHota$ & $\tieBTMaxAny$ \\
\bottomrule
\end{tabular}
\end{table}

Cases with $d > 0.002$ ($\Delta$HOTA/$\Delta$PTR, release minus frozen). v4: \tieShipBeyondCells{}. BT: \tieBTBeyondCells{}.
All are chute cases at ${\approx}0.6$ diameters/frame or more.

\emph{Two bug fixes made before release.} Every CBTV number in the paper comes from the release tree, which adds two fixes to the stable-sort tree.
(i) \emph{Frame-0 radius estimate.} CBTV sets its radius window from the particle radii at the first frame.
It previously estimated them with its own Laplacian-of-Gaussian detector, whose CPU and GPU implementations differ, so CPU and GPU runs could choose different windows; it now uses the radii of the shared frame-0 detections (rescaled to the detector's radius convention), so the decision no longer depends on the device.
In the corpus this changes the window on one flow only, the doubled-resolution large chute (6--11 px $\rightarrow$ 5--10 px).
(ii) \emph{Exit handling at the frame edge} (BT only; the v4 configuration does not use it). Tracks were terminated when their \emph{predicted} position fell outside the frame, which also killed particles resting at, or detected just past, the edge; only tracks that are leaving are now terminated.
Against the stable-sort values, BT is unchanged ($d < 5\times10^{-5}$) on $\fixBTExact$ of \ablNCells{} cases and within $0.002$ on $\fixBTWithin$ (beyond: \fixBTBeyondCells{}); v4 on $\fixShipExact$ and $\fixShipWithin$ (beyond: \fixShipBeyondCells{}).
The held-out cases are unchanged (max $d = \fixHeldMaxAny$).
No verdict, regime assignment, or ranking in the paper changes.

\emph{Selection rule re-checked.} With both configurations from the release tree, BT is within $0.002$ of v4 on both HOTA and PTR on every one of the $\eligNonFloor$ non-floor cases (v4 HOTA $\geq 0.15$; $\eligViolations$ violations), so BT remains \eligVerdict{} under the pre-registered rule.
\emph{Held-out rule re-checked.} On the release tree the held-out rule again returns \hoTzVerdict{} ($\hoTzNPass$ of \hoNCells{} cases; mean HOTA $\hoTzBTMean$ vs $\hoTzShipMean$).

\section{Supplementary: sensitivity of BT to its own parameters}
\label{sec:si:sensitivity}
Varying each parameter of the released configuration (BT) one at a time by a factor of two (the coast-velocity weight, the exit margin, and the colour-memory weight) changes HOTA by at most $\senMaxHota$ and PTR by at most $\senMaxPtr$ on average over \senNCells{} smooth and collision cases.
The \senMostSensitive{} is the most sensitive parameter; the largest change on any single case, $\senRbfkMaxHota$ HOTA, is on the doubled-resolution large chute at step~8.
Each BT parameter that has a scale was changed alone, to half and to twice its default, and the released configuration was re-run on the \senNCells{} smooth and collision cases of the corpus.
Table~\ref{tab:si:sensitivity} gives the change against the unmodified released configuration run on the same machine.
These runs are on Linux CPU, so the baseline is the same-machine re-run of BT, not the Windows GPU numbers of the main tables, and a difference measures the parameter alone.
The coast-velocity weight was only halved: its range is $(0,1]$ and its default is $1$, so doubling is not a valid value.
The defaults were not changed by this study.
The runs were made just before the frame-0 radius fix of Appendix~\ref{sec:si:tieorder}; the only cases that fix touches are on the doubled-resolution large chute, and those runs were repeated on the release tree with byte-identical tracks, so every number here holds for the release.
``Standing'' is the number of cases whose lead / tie / behind classification against the best other tracker (HOTA and PTR respectively, tie band $0.005$) changes.

\begin{table}[H]
\centering
\small
\caption{BT parameter sensitivity, one parameter at a time, \senNCells{} cases. Mean over cases of $\Delta = $ variant $-$ baseline; largest $|\Delta|$ with its case; number of cases whose standing against the best other tracker changes (HOTA/PTR).}
\label{tab:si:sensitivity}
\fitwidth{%
\begin{tabular}{@{}lrrrrr@{}}
\toprule
parameter & \senCoastHalfName{} & \senExitHalfName{} & \senExitTwoName{} & \senColHalfName{} & \senColTwoName{} \\
default $\rightarrow$ value & \senCoastHalfValue{} & \senExitHalfValue{} & \senExitTwoValue{} & \senColHalfValue{} & \senColTwoValue{} \\
\midrule
mean $\Delta$HOTA & \senCoastHalfMeanHota{} & \senExitHalfMeanHota{} & \senExitTwoMeanHota{} & \senColHalfMeanHota{} & \senColTwoMeanHota{} \\
mean $\Delta$PTR & \senCoastHalfMeanPtr{} & \senExitHalfMeanPtr{} & \senExitTwoMeanPtr{} & \senColHalfMeanPtr{} & \senColTwoMeanPtr{} \\
mean $|\Delta\mathrm{HOTA}|$ & $\senCoastHalfMeanAbsHota$ & $\senExitHalfMeanAbsHota$ & $\senExitTwoMeanAbsHota$ & $\senColHalfMeanAbsHota$ & $\senColTwoMeanAbsHota$ \\
max $|\Delta\mathrm{HOTA}|$ & $\senCoastHalfMaxHota$ & $\senExitHalfMaxHota$ & $\senExitTwoMaxHota$ & $\senColHalfMaxHota$ & $\senColTwoMaxHota$ \\
\quad case & \senCoastHalfMaxHotaCell{} & \senExitHalfMaxHotaCell{} & \senExitTwoMaxHotaCell{} & \senColHalfMaxHotaCell{} & \senColTwoMaxHotaCell{} \\
max $|\Delta\mathrm{PTR}|$ & $\senCoastHalfMaxPtr$ & $\senExitHalfMaxPtr$ & $\senExitTwoMaxPtr$ & $\senColHalfMaxPtr$ & $\senColTwoMaxPtr$ \\
\quad case & \senCoastHalfMaxPtrCell{} & \senExitHalfMaxPtrCell{} & \senExitTwoMaxPtrCell{} & \senColHalfMaxPtrCell{} & \senColTwoMaxPtrCell{} \\
standing changes (HOTA/PTR) & $\senCoastHalfStandHota$/$\senCoastHalfStandPtr$ & $\senExitHalfStandHota$/$\senExitHalfStandPtr$ & $\senExitTwoStandHota$/$\senExitTwoStandPtr$ & $\senColHalfStandHota$/$\senColHalfStandPtr$ & $\senColTwoStandHota$/$\senColTwoStandPtr$ \\
\bottomrule
\end{tabular}}
\end{table}

The mean absolute change is at most $\senMaxHota$ HOTA and $\senMaxPtr$ PTR for every variant; excluding the colour-memory weight it is at most $\senMaxHotaNoColour$ and $\senMaxPtrNoColour$.
The largest change on any case is on the doubled-resolution large chute at step~8, a case on which BT is close to failing on that machine ($\senRbfkBaseHota$ HOTA, $\senRbfkBasePtr$ PTR in the baseline run, against $\senRbfkMainHota$ and $\senRbfkMainPtr$ in the main tables; an unexplained difference seen only in one external environment outside the pinned environment): halving the colour-memory weight changes it by \senColHalfRbfkHota{} HOTA and \senColHalfRbfkPtr{} PTR.
Without that case no variant changes any single case by more than $\senMaxHotaNoRbfk$ HOTA or $\senMaxPtrNoRbfk$ PTR.

\section{Supplementary: record of the pre-registered selection}
\label{sec:si:prereg}
The released configuration (BT) was chosen from candidate configurations by rules written down before the corresponding runs, and the in/outflow questions were pre-registered before any of their runs.
The rules, the verdicts, the full per-case tables and the per-case values behind the numbers below will be archived with the data.

\emph{In/outflow questions (with Amendment~1).} Question~1 (no harm): the in/outflow configuration must stay within $0.002$ of v4 on HOTA and PTR on every one of the $\nCellsPrereg$ cases.
It failed on $\wsfQOneNFail$ cases, all on HOTA only, by $\wsfQOneDropMin$ to $\wsfQOneDropMax$, and $\wsfQOneNFloor$ of them are floor cases (v4 HOTA $< 0.15$).
Question~2 (edge benefit): a relative improvement of at least $25\%$ in a primary edge metric (birth-latency 90th percentile, exit-capture swaps, edge fragmentation) on at least $3$ of the $\wsfQTwoNCells$ registered edge cases, each with at least $30$ entry events.
It was met on $\wsfQTwoNMeet$ of $\wsfQTwoNCells$ cases.
Exit-capture swaps were $\wsfQTwoSwapsCells$ on every corpus and crop case in every configuration and $\wsfQTwoSwapsSynth$ on the synthetic in/outflow video, so the corpus does not show the failure the flags target.
Both are reported as negatives.

\emph{Candidate screen.} Rule fixed before the runs: a candidate is eligible if no case falls more than $0.002$ below v4 on HOTA or PTR; among eligible candidates the one with the highest mean HOTA over the $\nCellsPrereg$ cases is taken, ties to the fewest flags.
Table~\ref{tab:si:prereg} gives the outcome; BT is the only eligible candidate.
CT has a higher mean HOTA than BT but is ineligible under the rule.
CT adds a fourth mechanism, requiring two consecutive detections before a new track is confirmed; it lowered HOTA on the strip view (at step~1, from $\confKThreeHotaOff$ to $\confKThreeHotaOn$), raised the 90th-percentile delay before a new particle receives a track from $\confKThreeLatOff$ to $\confKThreeLatOn$ frames, and failed the selection rule on $\scrCtFail$ cases.
The selected configuration, BT, is labelled a post-hoc discovery on the evaluation corpus and was confirmed on new held-out flows (Section~\ref{sec:results:heldout}).

\begin{table}[H]
\centering
\caption{Candidate screen, all $\nCellsPrereg$ cases. v4 mean HOTA $\scrShipMean$. B = v4 + flow-prior coasting + exit kill; BT = B + track-memory colour; CT = BT + confirm-hits; C = B + confirm-hits.}
\label{tab:si:prereg}
\begin{tabular}{@{}lccc@{}}
\toprule
candidate & cases below v4 $-0.002$ & mean HOTA & eligible \\
\midrule
B & $\scrBFail$ & $\scrBMean$ & no \\
BT (released) & $\scrBtFail$ & $\scrBtMean$ & yes \\
CT & $\scrCtFail$ & $\scrCtMean$ & no \\
C & $\scrCFail$ & $\scrCMean$ & no \\
\bottomrule
\end{tabular}
\end{table}

\emph{Second adoption rule.} Before any run, a variant of BT was fixed as the one candidate replacement (BTN): its motion prior is switched from the dense PIV field to a nearest-velocity prior in the style of trackpy's \texttt{NearestVelocityPredict} (\texttt{--mu-source nvp}), in which each track's predicted displacement is its own last velocity when known and otherwise the median velocity of its nearest track with known velocity.
BTN was to replace BT only if it left no non-floor case below v4 $-0.002$ and raised the mean HOTA by at least $0.005$.
It left $\btnrNonFloorFail$ non-floor cases below that line and its mean HOTA is $\btnrMean$ against $\scrBtMean$ for BT, so BT stayed.

\section{Supplementary: local motion rigidity (pre-registered test)}
\label{sec:si:rigidity}
The rigidity explanation for where colour relaxation matches or exceeds CBTV was tested by a rule written down before any rigidity value was computed.
For each case, at the case's frame step and from ground truth only, $R$ is the fraction of pairs of particles within the relaxation tracker's neighbour radius ($3$ diameters) whose displacements over one step differ by less than its consistency tolerance ($0.35$ diameters), averaged over frames.
The outcome is the HOTA of the better relaxation variant minus that of CBTV (BT), over the $\rigN$ non-floor cases.
Because $R$ falls with displacement (Spearman $\rigRdRho$), the primary test is the partial Spearman correlation of $R$ with the outcome controlling for $\log_{10}$ displacement, one-sided.

\emph{Result.} On its pre-registered set of $\nCellsPrereg$ cases, the primary test passes: partial $\rho = \rigRho$, one-sided $p = \rigP$ ($n = \rigN$), so the pre-registered verdict is \emph{supported}.
On the corpus presented here the correlation is weaker and not significant (partial $\rho = \rigVElevenRho$, one-sided $p = \rigVElevenP$, $n = \rigVElevenN$).
The raw Spearman correlation is $\rigRawRho$.
The secondary prediction that all $\rigTercileOf$ collision cases fall in the lowest rigidity tercile is not met ($\rigTercileIn$ of $\rigTercileOf$).

\emph{Caveats.} Most cases sit near the rigidity ceiling (median $R = \rigRMedian$; $R \geq 0.97$ on $\rigNHigh$ of $\rigN$ cases), so the correlation is carried by the less rigid cases, which are mostly at high displacement.
Relaxation scores below CBTV on $\rigMNeg$ cases, above it on $\rigMPos$, and equal on $\rigMZero$.
One implementation bug was fixed before any statistic was read: on the strip-view videos of the pre-registered set the first ground-truth frames contain no mobile particles, so the particle diameter used to scale $R$ was undefined and six cases had no value; the diameter is now taken from the first frame that contains mobile particles, with the definitions unchanged.

\section{Supplementary: uncertainty of the benchmark numbers}
\label{sec:si:ci}
Confidence intervals come from a moving-block bootstrap over processed frames (blocks of $\ciBlock$ frames, $\ciB$ resamples, seed $\ciSeed$, $\ciLevel\%$ percentile intervals), applied to the stored tracks of every tracker and case on the benchmark branch and re-scored with the unmodified evaluator.
Significance is judged on \emph{paired} differences: in each resample the same frames are drawn for CBTV (BT) and for the best other tracker of that case, and the interval of the difference is reported.
We report absolute intervals for PTR but not for HOTA.
PTR is a ratio of sums over transitions and its bootstrap is unbiased (median $|$bias$|$ $\ciBiasPtrMedAbs$).
HOTA is not: resampling frames with replacement leaves each particle covered by fewer distinct frames, which inflates its association term, so the bootstrap distribution of CBTV's HOTA sits above the point estimate (median bias $\ciBiasHotaMed$, maximum $\ciBiasHotaMax$), and the point lies outside its own percentile interval in $\ciHotaOutside$ of $\ciHotaCells$ cases.
The bias cancels in the paired difference (median $|$bias$|$ $\ciBiasPairedMedAbs$), which is why only paired HOTA intervals are used.
The best other tracker is the maximum over all other tracker configurations in each case, chosen after the fact; counts of cases where CBTV is behind it are therefore upper bounds.

\begin{table}[H]
\centering
\caption{CBTV (BT) minus the best other tracker, by descriptive flow group (Section~\ref{sec:validation:regimes}): cases where the paired 95\% CI lies above 0 (ahead), below 0 (behind), or spans 0.}
\label{tab:si:ci}
\begin{tabular}{@{}lcccccc@{}}
\toprule
 & \multicolumn{3}{c}{HOTA} & \multicolumn{3}{c}{PTR} \\
flow group & ahead & behind & spans 0 & ahead & behind & spans 0 \\
\midrule
Impulsive collision & $\ciHotaAheadColl$ & $\ciHotaBehindColl$ & $\ciHotaNSColl$ & $\ciPtrAheadColl$ & $\ciPtrBehindColl$ & $\ciPtrNSColl$ \\
Agitated shear & $\ciHotaAheadAg$ & $\ciHotaBehindAg$ & $\ciHotaNSAg$ & $\ciPtrAheadAg$ & $\ciPtrBehindAg$ & $\ciPtrNSAg$ \\
Strip view and cylinder flow, step 8 & $\ciHotaAheadCrowd$ & $\ciHotaBehindCrowd$ & $\ciHotaNSCrowd$ & $\ciPtrAheadCrowd$ & $\ciPtrBehindCrowd$ & $\ciPtrNSCrowd$ \\
Strip view, swirl and cylinder flow, other steps & $\ciHotaAheadBeds$ & $\ciHotaBehindBeds$ & $\ciHotaNSBeds$ & $\ciPtrAheadBeds$ & $\ciPtrBehindBeds$ & $\ciPtrNSBeds$ \\
Smooth and structured chutes & $\ciHotaAheadSmooth$ & $\ciHotaBehindSmooth$ & $\ciHotaNSSmooth$ & $\ciPtrAheadSmooth$ & $\ciPtrBehindSmooth$ & $\ciPtrNSSmooth$ \\
Extreme displacement & $\ciHotaAheadExt$ & $\ciHotaBehindExt$ & $\ciHotaNSExt$ & $\ciPtrAheadExt$ & $\ciPtrBehindExt$ & $\ciPtrNSExt$ \\
Rotating drums & $\ciHotaAheadDrum$ & $\ciHotaBehindDrum$ & $\ciHotaNSDrum$ & $\ciPtrAheadDrum$ & $\ciPtrBehindDrum$ & $\ciPtrNSDrum$ \\
\midrule
all cases & $\ciHotaAhead$ & $\ciHotaBehind$ & $\ciHotaNS$ & $\ciPtrAhead$ & $\ciPtrBehind$ & $\ciPtrNS$ \\
\bottomrule
\end{tabular}
\end{table}

\end{document}